# THE H II REGION/PDR CONNECTION: SELF-CONSISTENT CALCULATIONS OF PHYSICAL CONDITIONS IN STAR-FORMING REGIONS

N. P. Abel,[1] G. J. Ferland,[1] G. Shaw,[1] P. A. M. van Hoof[2]

[1]University of Kentucky, Department of Physics & Astronomy, Lexington, KY 40506; npabel2@uky.edu, gary@pa.uky.edu, gshaw@pa.uky.edu

[2]Royal Observatory of Belgium, Ringlaan 3, B-1180 Brussels, Belgium
p.vanhoof@oma.be

## Abstract

We have performed a series of calculations designed to reproduce infrared diagnostics used to determine physical conditions in star forming regions. We self-consistently calculate the thermal and chemical structure of an H II region and photodissociation region (PDR) that are in pressure equilibrium. This differs from previous work, which used separate calculations for each gas phase. Our calculations span a wide range of stellar temperatures, gas densities, and ionization parameters. We describe improvements made to the spectral synthesis code Cloudy that made these calculations possible. These include the addition of a molecular network with ~1000 reactions involving 68 molecular species and improved treatment of the grain physics. Data from the Spitzer First Look Survey, along with other archives, are used to derive important physical characteristics of the H II region and PDR. These include stellar temperatures, electron densities, ionization parameters, UV radiation flux ($G_0$), and PDR density. Finally, we calculate the contribution of the H II region to PDR emission line diagnostics, which allows for a more accurate determination of physical conditions in the PDR.

## 1 Introduction

A major goal of infrared missions such as Spitzer, SOFIA, Hershel, and JWST is to understand the physical processes in galaxies and local regions of star formation. Such regions include starburst galaxies, Ultraluminous Infrared Galaxies (ULIRGs), embedded O stars in molecular clouds, and blister H II regions. Infrared spectroscopy can reveal conditions in ionized, photodissociated, and molecular gas. Observations combined with theoretical calculations reveal the temperature of the ionizing stars, $T_*$, electron density $n_e$, ionization parameter $U$, total hydrogen density $n_H$, and UV radiation field $G_0$. These, in turn, can yield information about the ionized, atomic, and molecular mass, the star formation rate, the chemical and thermal structure, elemental abundances, area and volume filling factors, number of stars required to produce the ionization, and the number of clouds and their radii (Wolfire, Tielens, & Hollenbach, 1990).

Up to now H II regions and PDRs have been treated as separate regions. This paper presents unified H II region / PDR calculations that make interpretation of observations on a self-consistent basis possible.

## 2 Physical Picture

### 2.1 Structure of Clouds with Embedded Ionizing Stars

Evans (1999) outlines the structure of molecular clouds, or star-forming regions. As gravitational collapse of the cloud takes place, O stars hot enough to ionize H create an H II region. The level of ionization of an H II region can be characterized by $U$, the dimensionless ratio of the flux of ionizing photons, $\phi_H$, to $n_H$ at the illuminated face, given by:

$$U = \frac{\phi_H}{n_H c} \tag{1}$$

where $c$ is the speed of light. Typical values for $U$ vary from about $10^{-1}$ to $10^{-4}$ (Veilleux & Osterbrock 1987).

These O stars are either embedded deep in the molecular cloud or form on the edge of the molecular cloud and create a blister-type H II Region, similar to the Orion Nebula (O'Dell, 2001). In the case of starbursts, the same physical mechanisms probably occur, only on a much larger scale.

Hydrogen makes the transition from ionized to atomic form when all photons with energies $h\nu > 13.6$ eV are absorbed. Once hydrogen becomes primarily atomic, the H II region ends and the Photodissociation Region, or PDR, begins (Tielens & Hollenbach, 1985). Hydrogen makes the transition from atomic to molecular across the PDR. Other molecules can form after $H_2$ since $H_2$ is a primary catalyst.

PDRs are characterized by $n_H$ and the strength of the Far-Ultraviolet (FUV) field ($G_0$) between 6 and 13.6 eV relative to the background interstellar radiation field (Habing 1968).

$$G_0 = \frac{\int_{912\text{Å}}^{2067\text{Å}} I_\lambda d\lambda}{1.6 x 10^{-3} \text{ (ergs cm}^{-2}\text{ s}^{-1})} \tag{2}$$

The value of $G_0$ strongly influences the physical conditions in PDRs (Wolfire, Tielens, & Hollenbach, 1990; Kaufman et al., 1999). Hollenbach & Tielens (1997) give a complete description of the physical processes in a PDR.

## 2.2 Infrared Diagnostics – Theory and Observations

### 2.2.1 H II Region Diagnostics

$T_*$ and $U$ are determined through observations of lines of more than one ionization stage of the same element. For instance, a common diagnostic used in H II regions is the ratio of [Ne III] 15.56 μm to [Ne II] 12.81 μm (Verma et al., 2003; Giveon et al., 2002). A higher line ratio corresponds to a higher level of ionization, corresponding to a higher $T_*$ and $U$. Other line ratios that can characterize the radiation field are [S IV] 10.51 μm to [S III] 18.71 μm or 33.48 μm; [N III] 57.21 μm to [N II] 121.7 μm or 205.4 μm line (Rubin et al, 1994); and the ratio of [Ar III] 8.99 μm to [Ar II] 6.99 μm.

The electron density is determined through ratios of lines of the same ion (Osterbrock & Ferland 2005). The average electron energy is much greater than the excitation potential for mid-infrared transitions, eliminating the temperature dependence of the line ratio. However, each line will have a different critical density, which makes the line ratio dependent on $n_e$. Examples of density diagnostics in the infrared are the [O III] 52 μm line to [O III] 88 μm, [S III] 18.71 μm to [S III] 33.49 μm, and [N II] 121.7 μm to [N II] 205.4 μm (Rubin et al. 1994; Malhorta et al. 2001).

### 2.2.2 PDR diagnostics

Infrared fine-structure lines can also determine the physical conditions in the PDR and molecular cloud. These lines include [C II] 158 μm, [C I] 370 μm, 609 μm, [O I] 63 μm, 146 μm, the [Si II] 35 μm, and [Fe II] 1.26μm, 1.64μm lines. These are important coolants in PDRs (Hollenbach & Tielens, 1997). Observations of the intensity of these lines combined with theoretical calculations help to determine physical conditions in star forming regions (Tielens & Hollenbach 1985; Wolfire, Tielens, & Hollenbach 1990). Kaufman et al.'s (1999) calculations show how these lines vary over a wide range of $n_H$ and $G_0$. The basic premise is that, by observing as many lines as possible, theoretical calculations can be compared to observations to determine $n_H$ and $G_0$. These then determine such properties of star forming regions as the mass of the cloud, number of clouds, filling factor, chemical composition, and temperature.

### 2.2.3 H II Region Contributions to PDR Diagnostics

Previous calculations have treated the H II region and PDR separately. PDR calculations typically do not consider hydrogen-ionizing radiation, and define a PDR as starting where H is predominantly atomic. Diagnostic emission lines are assumed to originate in only one of these regions.

Some ions (for example, $C^+$ or $Si^+$) are present in both the H II region and PDR, and their lines, in this case [C II] 158 μm and [Si II] 35 μm, can form in both. Even the [O I] 63 μm, 146 μm lines can form in H II regions (Aannestad & Emery, 2003). Calculations that ignore the H II region (essentially all current PDR simulations) will underestimate the line intensity, leading to incorrect value of $n_H$ and $G_0$. This can lead to erroneous conclusions for the physical conditions.

It is already known that the H II region can contribute to "PDR" lines. Kaufman et al. (1999) note the need to correct observations for any possible H II region emission before using their contour plots. There are several ways to make this correction, both observational and computational. Petuchowski & Bennett (1993) and Heiles (1994) note that the contribution of the H II region intensity to the [C II] 158 μm line should be proportional to [N II] 122 μm. The most common computational approach is to run a separate model of the H II region to estimate its contribution to a PDR diagnostic (Carral et al. 1994). They found that as much as 30% of the [C II] line can come from the H II region, with the greatest contribution coming from low-density H II regions adjacent to PDRs. A study of S125 by Aannestad & Emery (2003) found that ~40% of the [C II] 158 μm and ~20% of the [O I] 63 μm line intensity came from the H II region.

The overall approach of considering the two regions separately has many difficulties. One is that it is nearly impossible to match boundary conditions for two separate simulations. The density of the PDR, relative to the H II region, is assumed to be the inverse ratio of their temperatures, $\approx 10^2$, as suggested by gas pressure equilibrium. The transition between the H II region and PDR is not included, leading to a lack of continuity.

Dynamical processes link these two regions in nature (see Henney et al. 2005). The physical properties of the PDR are a consequence of the transport of gas and radiation through the H II region, while the H II region is the result of the flow of gas through the PDR. Our approach is to compute self-consistently the temperature, density, and ionization starting with the H II region and progressing through the PDR.

We discuss computational details in Section 3 and in Appendices A, B, and C. In Section 4 we describe the physical processes for a standard model that begins in the H II region and extends deep into the molecular cloud. Section 5 shows how H II region and PDR infrared diagnostic lines vary with $U$, $n_{H^+}$, and $T_*$, under the assumption of constant pressure. We will also determine the H II region contribution to commonly used PDR diagnostic emission lines. We will then show applications to our calculations in Section 6, and end with a list of conclusions in Section 7.

## 3 Calculations

We use the developmental version of the spectral synthesis code Cloudy, last described by Ferland et al. (1998). Shaw et al. (2005) and Henney et al. (2005) describe recent advances in its treatment of $H_2$ and dynamics.

We have expanded the molecular reaction network to include ~1000 reactions involving 68 molecules. We describe the molecular network in Appendix A and in Abel et al. (2004). Our chemical equilibrium agrees very well with other PDR codes presented at the Leiden 2004 PDR workshop (http://hera.ph1.uni-koeln.de/~roellig/).

Van Hoof et al. (2004) describe the grain physics. We self-consistently determine the grain temperature and charge as a function of grain size and material, for the local physical conditions and radiation field. This determines the grain photoelectric heating of the gas, an important gas heating process, as well as collisional energy exchange between the gas and dust. The rates at which $H_2$ forms on grain surfaces is derived using the temperature- and material-dependent rates given in Cazaux & Tielens (2002). We include grain charge transfer as a general ionization – recombination process, as described in Appendix B. We also treat stochastic heating of grains as outlined in Guhathakurtha & Draine (1989), which can affect the dust continuum shape.

Primary and secondary cosmic ray ionization processes are treated as described in Appendix C. We include a rigid-rotor model for the $v = 0$ rotation levels of CO. This predicts line intensities for the lowest 30 rotational transitions.

Our calculations include all stages of ionization for the lightest 30 elements. We use a complete model of the hydrogen atom in calculating the H I emission-line spectrum and ionization rates (Ferguson & Ferland 1997), as well as a complete model of the helium isoelectronic sequence (Porter et al. 2005). Ionization processes include direct photoionization, charge transfer, the Auger effect, dielectronic, collisional, and cosmic ray ionization. In the H II region, the most important of these processes is photoionization, although charge transfer ionization can be important for some species, in particular oxygen (Kingdon and Ferland 1999).

We cover the range of observed H II regions, with $n_{H^+}$ between $10^1$ and $10^4$ cm$^{-3}$, and $U$ between $10^{-4}$ and $10^{-1}$. Both $n_{H^+}$ and $U$ were incremented by 1 dex. All calculations stop at an $A_V$ of 100 magnitudes, ensuring that they extend deep into the molecular cloud.

We use the non-LTE CoStar stellar atmospheres (Schaerer & de Koter, 1997) and vary the $T_*$ from 34,000 to 46,000 K in steps of 4,000 K. This range of $T_*$ covers stars roughly ranging from spectral type O8 to O5 (Vacca et al. 1996), and spans the calculations performed by Giveon et al. (2002). Combining this range of $T_*$ with the range of $n_{H^+}$ and $U$, we have a total of 64 calculations.

Our calculations depend sensitively on the stellar continuum. Figure 1 shows continua emitted by single stars of temperatures 33,500 K, 35,000 K, 40,000 K, and 48,000 K, roughly corresponding to O8 to O5 stars. We assumed the temperature-luminosity relation given by Vacca et al. (1996). As $T_*$ increases, the total luminosity of the star also increases. However, the Lyman continuum luminosity, which sets the scale of the H II region, increases far more rapidly than the luminosity of the Balmer continuum, which activates the PDR. This means that, for a fixed density, the size of the H II region relative to the PDR increases with increasing $T_*$. As we discuss in Section 5.3, this has important consequences to the interpretation of PDR emission-line diagnostics.

We use gas-phase abundances based on an average of the abundances in the Orion Nebula, with the complete set of abundances given in Baldwin et al. (1996). For the most important species, the abundances by number are He/H = 0.095, C/H= $3\times10^{-4}$; O/H= $4\times10^{-4}$, N/H= $7\times10^{-5}$, Ne/H= $6\times10^{-5}$, and Ar/H= $3\times10^{-6}$. We have assumed S/H= $2\times10^{-6}$ based on observations of starburst galaxies by Verma et al. (2003). The complete set of assumed fractional abundances can be found in Ferland (2002).

We assume a galactic ratio of visual extinction to hydrogen column density, $A_V/N(H)$, of $5\times10^{-22}$ mag cm². Grain size distributions for gas adjacent to H II regions, such as Orion (Cardelli et al. 1989) and starburst galaxies (Calzetti et al. 2000) tend to have a larger ratio of total to selective extinction than observed in the ISM. We therefore use a truncated MRN grain size distribution (Mathis et al. 1977) with $R = 5.5$, which reproduces the Orion extinction curve (Baldwin et al. 1991).

We also include size-resolved PAHs in our calculations, with the same size distribution that is used by Bakes & Tielens (1994). The abundance of carbon atoms in PAHs that we use, $n_C(PAH)/n_H$, is $3\times10^{-6}$. PAHs are thought to exist mainly in regions where hydrogen is atomic (Giard et al. 1994). We therefore scale the PAH abundance by the ratio of $H^0/H_{tot}$ ($n_C(PAH)/n_H = 3\times10^{-6}\times [n(H^0)/n(H_{tot})]$).

We link the H II region and PDR by assuming constant total pressure. This equation of state includes radiation pressure, both from the stellar continuum and internally generated light, gas pressure, and turbulent, ram, and magnetic pressures when appropriate (Baldwin et al. 1991; Henney et al. 2005). Constant pressure is a first approximation to the actual flow (see Henney et al. 2005) and was assumed by previous authors (Carral et al. 1994).

We will distinguish between the H II region and PDR contributions to an emission line in the following. We define the start of the PDR as the position where $n(H^+)/n(H_{tot})$ falls below 1% (Shields & Kennicutt 1995). Tests show that the relative contributions are not sensitive to smaller values of this threshold, while larger values will diminish the H II region contribution.

# 4 A single cloud

In order to calculate the physical conditions in the ionized, atomic, and molecular gas, we must include all relevant physical processes in each regime. In this section we describe the chemical and thermal structure of a typical cloud. We also describe important physical processes and their treatment. We show in Figures 2-15 the results of our calculations for $T_* = 38{,}000$ K, $U = 10^{-2}$, and $n_{H^+}$ of 3,000 cm$^{-3}$. Other parameters are the same as those given in Section 3.

## 4.1 Incident and Transmitted Continuum

Figure 2 shows the incident and transmitted continuum at three different points in our calculation. These points are where the ionized hydrogen fraction falls to less than 1%, the point where the $H_2$ fraction reaches 50%, and the shielded face of the cloud.

This plot shows the importance of including emission from the H II region in calculating the conditions in the PDR. The transmitted continuum at an ionization fraction of 1% is the radiation field that is incident upon the PDR, given our definitions of these regions. Photoelectric and dust absorption in the H II region removes the Lyman continuum photons. Additionally, dust re-emits absorbed starlight at infrared wavelengths. This causes a rise in the transmitted continuum for $\lambda > 1$ µm. Deeper in the cloud, molecular dissociation, grain absorption, and photoionization of species with ionization potentials less than 13.6 eV absorb the Balmer continuum. This continuum will depend strongly on the properties of the H II region, and this forms a further link between two regions. All radiation with wavelengths $\lambda < 1$ µm is absorbed at the shielded face of the cloud. This leaves only cosmic rays and the infrared spectrum to heat the gas.

Figures 3-7 show the ionization structure for H, O, C, Ne, and Si across this cloud. We use $A_V$ as our depth parameter, which is customary for PDR calculations (Tielens & Hollenbach 1985) but not for H II regions. For Figure 3, we show the density of $H^+$, $H^0$, and $H_2$ vs $A_V$, while Figures 4-7 show the ionization fraction, defined as $n(X^{i+})/n(X_{tot})$ vs $A_V$. Figure 3 shows how, under the assumption of constant pressure, $n_H$ increases as $T$ decreases. For reference, the depth into the H II region is given by $R-R_0 = 6.32\times10^{17} \times A_V$ (cm).

## 4.2 H II Region

The radiation field of the ionizing source produces an H II region at the illuminated face. Most of the heavy elements are doubly ionized due to the high $U$ ($10^{-1.5}$) and $T_*$ (38,000 K). Higher-energy photons are absorbed as we approach the H ionization front.  Species with ionization potentials greater than hydrogen become atomic, while species with ionization potentials less than hydrogen remain either singly (ex. C, Si, S) or doubly (ex. Ca or Sc) ionized beyond the ionization front.

The ionization front occurs at an $A_V$ of ~1.  Classical Strömgren sphere theory (Osterbrock and Ferland 2005), predicts that the ionization front should occur at an $A_V$ of 1.5 for this ionization parameter and dust-to-gas ratio.  The size of the H II region is smaller due to absorption of a portion the Lyman continuum by dust (Bottorff et al. 1998).  Dust also strongly absorbs the L$\alpha$ line.  Rees, Netzer, & Ferland (1989) and Ferland et al. (1998) discuss the solution of the multi-level hydrogen atom and transport of resonance line radiation.

The temperature (Figures 8 and 9) is determined by energy conservation.  Figures 10-12 gives the total heating & cooling rates, and identifies important heating and cooling processes.  Since energy is conserved, total heating = total cooling.  We therefore only show a single rate in Figure 10.  The H II region temperature is ~$10^4$ K and the dominant heating / cooling processes are photoionization of hydrogen balanced by fine-structure excitation of optical and infrared lines of [O II], [O III], and [N II] .  The single most important coolant is the [O III] 5007Å line.

## 4.3 PDR

The transmitted radiation from the H II region (Figure 2) determines the conditions in the PDR.  Hydrogen makes the transition from atomic to molecular and other molecules form, as shown in Figure 3.

Grain physics dominates conditions in the PDR.  Grain photoelectric heating is the dominant heating mechanism (Figure 11), although photoionization of carbon is significant.  PAHs have a large abundance in the atomic hydrogen region, and these small grains contribute significantly to the total heating (Bakes & Tielens 1994), although graphite and silicate grains are also significant.  As Figure 9 shows, the dust temperature depends on grain type and size (van Hoof et al. 2004).  Dust also extinguishes the radiation field across the PDR.

Figure 13 shows the cumulative emission in each line integrated from the illuminated face to any depth in the cloud, for [O I] 63 μm, [O I] 146 μm, [C II] 158 μm, and [Si II] 35 μm.  These lines are predominantly excited by collisions with $H^0$.  Almost all emission in these four lines comes from the PDR, with a negligible contribution from the H II region.  Figure 12 shows that these lines are also the dominant coolants in the PDR.  In dense PDRs, these lines can heat the gas when radiative excitation followed by collisional de-excitation becomes important (see Figure 11; also Tielens & Hollenbach 1985).

Line and dust heating is sensitive to the continuum transmitted through the H II region.  Dust absorbs the UV continuum and re-emits it in the infrared (Figure 2).  This H II region continuum enters the PDR and some is absorbed by lines and dust, eventually heating the gas. The dust continuum emitted by the H II region must be self-consistently treated if we are to correctly treat line and dust heating, and other processes that depend on the IR continuum. Our calculations self-consistently treat these physical processes, since we calculate the dust continuum produced by the H II region and its effects on the PDR.

The dust temperatures influence molecular formation, since they are important in determining the molecular reaction rates, in particular the $H_2$ formation rate on grain surfaces (we use Cazaux & Tielens 2002). In addition, the FUV radiation field (11.2 eV ≤ hν ≤ 13.6 eV) destroys $H_2$ by the Solomon process. Our calculations can include a detailed microphysical treatment of $H_2$ to treat the Solomon process (Shaw et al. 2005), although for this calculation we treat the $H_2$ assuming the equivalent three-level molecule (Tielens & Hollenbach 1985) and treat dissociation using the self-shielding function of Draine & Bertoldi (1996). $H_2$ forms quickly once all photons greater than 11.2 eV are extinguished. This occurs at about an $A_V$ of ~4.5.

Other molecules form once $H_2$ forms (Figure 14), since it acts as a catalyst for the formation of other molecules through a series of ion-molecule reactions (Prasad & Huntress 1980). A complete description of the molecules considered in our reaction network is given in Appendix A. The second-most abundant molecule that forms after $H_2$ is CO. Shortly after $H_2$ forms, carbon makes the transformation from $C^+$ to CO (with a small amount of C) at an $A_V$ of ~8. Figure 13 also shows the integrated emission in the lowest three rotational transitions for CO.

## 4.4 Molecular Cloud

The gas is predominantly molecular deep in the cloud. CO forms at the point where its electronic bands are well shielded and other molecules form in deeper regions. The most important are $N_2$, $O_2$, $H_2O$, OH, SO, and CN. All molecules reach their asymptotic abundances by an $A_V$ of ~30.

CO collisional excitation is the primary coolant in the molecular gas (Figure 12). Both $^{12}CO$ and $^{13}CO$ are included in our calculation, and both contribute to the total cooling.

The gas is heated primarily by grain collisions. Figure 15 a-b shows the grain opacity and the transmitted continuum in the 1-100 μm range at an $A_V$ of 20. The integral of these two quantities is proportional to the local heating of each grain species. In the 10-40 μm range, the silicate opacity is considerably higher than the graphite opacity. Absorption of the infrared continuum by silicate grains causes the silicates to be hotter than other gas constituents. In the molecular region the gas and graphite grains have $T \approx 45$ K while the silicates are typically 5 K hotter. Collisional energy exchange between these cooler constituents and the warmer silicates then heats the gas. In addition to grain collisions, line de-excitation and cosmic ray ionization makes a small contribution to the total heating.

## 5 Results and Discussion

### 5.1 Density

Here we present the results of our grid of calculations, spanning a wide range of $U$, $T_*$, and $n_{H^+}$. Here $n_{H^+}$ is just the ionized hydrogen density at the illuminated face, which for all calculations is also $n_H$ at the illuminated face.

Our results are presented as a series of contour plots (Figures 16 – 33). The value of $n_{H^+}$ at the illuminated face (and hence the initial value of $n_H$) is the x-axis on the contour plots. Figure 16 and Figure 17 shows the total hydrogen density at two important depths. Figure 16 shows $n_H$ at the point where the ionized hydrogen fraction falls to less than 1%. In general, $n_H$ at this point has increased by a factor of ~5-20, corresponding to the decrease in temperature from ~$10^4$ K in the ionized region to 500-2,000 K at the start of the PDR. Figure 17 shows $n_H$ at a deeper point, where the fraction of hydrogen in the form of $H_2$ exceeds 90%. Most PDR lines will have fully formed by this point, where $T \sim 100$ K and $n_H$ has increased by about two orders of magnitude. Although not plotted, there is an additional increase in $n_H$ in deeper regions as other molecules, such as CO, form and the gas becomes even colder.

Most PDR calculations assume a constant $n_H$, so no direct comparison is possible between the density in our calculations and the density used in PDR calculations. The densities shown in Figure 16 and Figure 17 are close to the lower and upper limits to $n_H$ that would be used in equivalent constant-density PDR calculations. For example, consider a case where $T_* = 38,000$ K, $n_H = 100$ cm$^{-3}$ and $U = 10^{-2}$. This set of parameters yields an $n_H$ at the face of the PDR of ~$2\times10^3$ cm$^{-3}$ and ~$2\times10^4$ cm$^{-3}$ when all hydrogen is in the form of $H_2$. The average density in the PDR falls somewhere in between.

### 5.2 H II Emission Line Diagnostics

In this section, we discuss our results for infrared emission lines that are produced in the ionized gas.

Figure 18 to 21 show our ratios involving different stages of ionization of the same element. We will use $R(X)_{denom}^{numer}$ as shorthand notation for all line ratios. Here X refers to the element, and *numer/denom* refers to the wavelengths of the upper and lower ionization stages. The four line ratios we calculate are $R(Ne)_{12.8\mu m}^{15.5\mu m}$, $R(S)_{10.5\mu m}^{18.7\mu m}$, $R(N)_{122\mu m}^{57\mu m}$, and $R(N)_{205\mu m}^{57\mu m}$. Observations of these emission-line ratios can determine $T_*$ and $U$. These two parameters in turn yield information on the total flux of ionizing photons, the number of stars required to produce the observed level of ionization, and the age of the starburst.

Each figure shows several general trends. Each ratio is sensitive to both $U$ and $T_*$. As $T_*$ increases, the incident continuum produces more high-energy photons, leading to greater production of Ne$^{2+}$, N$^{2+}$, and S$^{3+}$, which leads to stronger emission from these species. The results are very similar for a $T_*$ of 42,000 K and 46,000 K. In the case of the nitrogen lines, the critical density of the [N III] 57 μm line is ~$2.7 \times 10^2$ cm$^{-3}$, so collisional de-excitation effects become important for larger densities. This is not the case for the neon and sulphur lines, whose critical densities exceed $10^4$ cm$^{-3}$. Figure 18 and 19 show that, for the range of $n_{H^+}$ considered here, the Ne and S ratios depend only on $T_*$ and $U$.

Ratios of lines emitted by the same ion are sensitive to density rather then $T_*$ or $U$. Figure 22 shows three such density diagnostics, $R(S)_{33.5\mu m}^{18.7\mu m}$, $R(O)_{88\mu m}^{52\mu m}$, and $R(N)_{205\mu m}^{122\mu m}$, for $T_* = 38,000$ K. The contours are nearly vertical, confirming their utility as density diagnostics. Note that $n_{H^+}$ is, to within 10%, $n_e$ in the H II region.

## 5.3 PDR Emission Line Diagnostics

Calculated intensities (in units of erg cm$^{-2}$ s$^{-1}$) of common PDR diagnostic emission lines are shown in Figure 23-26. Shown are the predicted intensities of the [O I] 63 μm, [O I] 146 μm, [C II] 158 μm, and [Si II] 35 μm lines. Observations of these lines can be combined with theoretical calculations to determine $n_H$ and $G_0$ (Wolfire et al. 1991; Kaufman et al. 1999). This in turn helps to determine the mass of the molecular and atomic gas, number and radii of clouds, chemical and thermal structure, and the volume-filling factor. Figure 27 shows the dependence of $G_0$ on $U$ and $n_H$. $G_0$ scales linearly with both parameters because it is proportional to the flux. Additionally, $G_0$ does not depend strongly on $T_*$. Comparing Figures 23-26 with Figure 27 shows that the line intensities closely follow the UV flux, as expected. These figures give line intensities rather than line ratios, and are directly applicable for nearby resolved PDR regions. Line ratios must be used for observations of extragalactic sources, which are not spatially resolved. Figure 29 and 29 give the ratio of [O I] 63 μm to [O I] 146 μm, and the ratio of [O I] 63 μm to [C II] 158 μm. These include contributions from the ionized gas in addition to the PDR.

## 5.4 Ionized Gas Contributions to PDR Emission

We show in Figure 30-33 the percentage contribution of emission from the ionized gas to the PDR diagnostic lines shown in Figure 23-26.

The ratio $G_0/n_H$ and $T_*$ are the two most important factors that determine the contribution from ionized gas to PDR lines. The emission lines shown in Figure 23-26 are generally excited by H$^0$ collisions in the PDR. The ratio $G_0/n_H$ determines the size of the atomic region. For high values, $H_2$ is photodissociated to a larger depth and the size of the atomic hydrogen region is large, while $H_2$ forms quickly for small values. $T_*$ is important because it determines the continuum shape, and hence the ratio of the flux of UV photons to the flux of hydrogen-ionizing photons. For regions with a large H II region but a small atomic region, a larger fraction of the PDR line will form in the H II region. As a result, the predictions depend on the continuum shape.

By combining Figures 1, 27 with 30-33, we can understand the contribution from ionized gas to PDR emission lines. As $T_*$ increases, more hydrogen ionizing flux is generated relative to $G_0$ (see Figure 1 and 27). This increases the amount of ionized gas relative to the atomic gas. The relative increase in ionized to atomic gas causes more of the total PDR diagnostic line emission to come from the H II region. We find that for $T_* > 42,000$ K, combined with $G_0 < 100$, at least 10% of the emission from all four PDR diagnostic lines considered emerges from the ionized gas. For the [O I] lines there is a good correlation between the relative sizes of the H II region and the atomic region and the contribution to the [O I] lines from the ionized gas. The ionization structure of O closely follows that of H since the two are strongly coupled by charge transfer, and their ionization potentials are nearly equal. The contribution of the ionized gas is greater for the 146 μm line. The 63 μm line is always stronger in the PDR due to its lower excitation potential while the lines have similar intensities in the H II region.

The [Si II] 35 μm and [C II] 158 μm emission from the ionized gas is significant over a wider range of $U$, $n_{H^+}$, and $T_*$. The ionization potentials of atomic carbon and silicon are less than 13.6 eV so the first ions are present in both regions. At least 10% of the 35 μm emission and 5% of the 158 μm emission comes from the ionized gas. We still see the general trend that higher $T_*$ combined with a lower UV flux produces more total emission from the ionized gas.

We should note once again that our calculations all assume that the ionized, atomic, and molecular gas are in pressure equilibrium. The equation of state is dominated by gas pressure since turbulent and magnetic pressures are not included. If turbulent or magnetic pressure dominates then the density law will be different. We further assume a static geometry. Currently an effort is underway to incorporate dynamics into our calculations. This will allow us to treat the dynamical interaction of H II regions with molecular clouds. As explained in Bertoldi & Draine (1996), when advection is included, the ionization and molecular dissociation front can merge. The predicted size of the ionized and atomic regions will change, making the ionized region larger compared to the atomic region. As our calculations above show, this will lead to a larger contribution of the ionized gas to PDR diagnostics. This is an important future area of study to interpret fully the emission line spectra of star-forming regions.

# 6 Applications to Observations

Our calculations have many applications to astrophysical environments. Most importantly, we have outlined a methodology where we self-consistently calculate the emission-line spectrum of the ionized, atomic, and molecular gas along with continuum emission from the dust.

Observations of lines from different ions of the same element determine $T_*$ and $U$ (Figures 18-21), while line ratios from the same ion can determine the gas density (Figure 22). For our assumed equation of state (constant pressure), $T_*$, $U$, and $n_{H^+}$ also determine the conditions in the PDR. Consider an example. For $T_* = 38{,}000$ K, $U = 0.01$, and $n_{H^+} = 100$ cm$^{-3}$, we can use Figures 16, 17, and 27 to determine $n_H$ and $G_0$ for the atomic and molecular gas. Of course, we can also invert this logic. If one makes observations of PDR emission lines, we can determine $G_0$ and $n_H$ in the PDR. We can then use Figures 16, 17, and 27 to determine what values of $U$ and $n_{H^+}$ reproduce the observations.

We next apply our calculations to some of the early-published results of the Spitzer mission. Below are examples of extragalactic observations where we deduce some physical characteristic of the region of interest by comparing our calculations to the observed spectral features.

### *6.1.1 NGC 253*

NGC 253 is a starburst galaxy with a wealth of infrared observations. Previous works by Carral et al. (1994), Verma et al. (2003), and Devost et al. (2004) have deduced abundances, $T_*$ (34,500 $\pm$ 1,000) K, $Q$(H) ($10^{53.2}$ photons s$^{-1}$), $G_0$ (~$2\times10^4$), $n_e$ ($430^{+290}_{-225}$ cm$^{-3}$) and a density of the atomic gas of $10^4$ cm$^{-3}$. Additionally, Carral et al. (1994) find that the ionized and atomic regions are in gas pressure equilibrium.

Since NGC 253 is in gas pressure equilibrium, we can use our methodology to calculate the conditions in the ionized and atomic gas simultaneously. Comparing Figure 22 to the observed ratios $R(\mathrm{S})^{18.7\,\mu m}_{33.5\,\mu m}$ of 0.5 (Verma et al., 2003) and $R(\mathrm{O})^{52\,\mu m}_{88\,\mu m}$ of ~1-2 (Carral et al. 1994), we find $n_{H^+}$ of ~100-200 cm$^{-3}$, significantly lower than the previous results. The observed ratio $R(\mathrm{Ne})^{15.5\,\mu m}_{12.8\,\mu m}$ of 0.14 – 0.2 (Devost et al. 2004), together with their $T_*$ corresponds to $U$ ~0.01.

Let us now consider the PDR diagnostic lines. We will adopt 0.01 for $U$, 34,000 K for $T_*$, and 150 cm$^{-3}$ for $n_{H^+}$. Figure 29 shows that the predicted ratio of [O I] 63 µm to [C II] 158 µm (~1), for the derived $U$ and $n_{H^+}$, is in excellent agreement with the observed ratio of 0.8-1.1 (Carral et al. 1994). Our calculations suggest that $G_0$ is ~$5\times10^3$ (Figure 27), about four times lower than what Carral et al. (1994) deduce. Had we used $n_e$ = 430 cm$^{-3}$, as deduced by Carral et al. (1994) rather than the density derived from the [S III] lines, we would have obtained their $G_0$. However, this would then overestimate the [O I] to [C II] line ratio by a factor of 3. Using Figure 16 and 17, we estimate an $n_H$ in the PDR of 2,000 - 20,000 cm$^{-3}$. The density derived by Carral et al. (1994) of 10$^4$ cm$^{-3}$, lies within the range of densities we derive.

We wish to highlight two important points regarding our analysis of NGC 253. First, $T_*$ and two H II-region line ratios determine $U$ and $n_H$ in the ionized, atomic, and molecular gas. Calculations with these parameters reproduce the observed [O I] 63 μm to [C II] 158 μm PDR line ratio. This supports the conclusion by Carral et al. (1994) that NGC 253 is in gas pressure equilibrium. Secondly, for our values of $U$ and $n_{H^+}$, we find that ~30% of the [C II] 158 μm line and 20% of the [Si II] 35 μm line are produced in the ionized gas. Carral et al. (1994) found a similar estimate of the contribution of the H II region to the [C II] line. Had we ignored the H II region, we would have overestimated the [O I] 63 μm to [C II] 158 μm line ratio by a factor of ~1.4, resulting in a lower $G_0$ and $n_H$ in the PDR. This would have led to large uncertainties in conditions derived with these two parameters.

### *6.1.2 NGC 7331*

NGC 7331 is a nearby spiral galaxy observed by Spitzer as part of the SINGS survey (Smith et al. 2004). They observed $R(\mathrm{Ne})^{15.5\mu m}_{12.8\mu m}$ and $R(\mathrm{S})^{18.7\mu m}_{33.5\mu m}$, together with the lowest three rotational lines of the ground vibrational state of H$_2$. They derived densities in the ionized and atomic region of < 200 cm$^{-3}$ and 5000 cm$^{-3}$, respectively. They found $G_0$ ~10$^2$-10$^3$ based on the H$_2$ lines and the dust continuum, and conclude that the region is in pressure equilibrium. Following the same procedure as NGC 253 we compared our calculations with SST observations. According to Figure 22, the observed ratio $R(\mathrm{S})^{18.7\mu m}_{33.5\mu m}$ of 0.4-0.67 indicates $n_{H^+}$ < 200 cm$^{-3}$, in agreement with Smith et al. (2004). We therefore adopt a value of 150 cm$^{-3}$, which gives $R(\mathrm{S})^{18.7\mu m}_{33.5\mu m}$ =0.55. Unfortunately, we do not have an estimate of $T_*$. The observed $R(\mathrm{Ne})^{15.5\mu m}_{12.8\mu m}$ of 0.64–0.75 implies, from Figure 18, that $T_* > 34,000$ K. This is because $T_* \leq 34,000$ K requires that $U > 10^{-1}$, an unlikely result. Assuming Smith et al.'s (2004) $G_0$, Figure 27 combined with Figure 18 implies $T_* < 38,000$ K. We therefore assume $T_* = 36,000$ K ± 2,000 K. Assuming this value of $T_*$, we vary $U$ to reproduce the observed $R(\mathrm{Ne})^{15.5\mu m}_{12.8\mu m}$. Our results were: Log[$U$] = -2.5; $R(\mathrm{Ne})^{15.5\mu m}_{12.8\mu m}$ = 0.68 (0.64–0.75); $R(\mathrm{S})^{18.7\mu m}_{33.5\mu m}$ = 0.55 (0.40-0.67); $G_0$ = 9×10$^2$ , $n_H$ (PDR) = 3×10$^3$ cm$^{-3}$, and $n_H$ (H$_2$ fully forms) = 4×10$^4$ cm$^{-3}$. The values of $U$ and $T_*$ are new to this work, the other parameters are in agreement with Smith et al. (2004).

# 7 Conclusions

In this paper, we have done the following:

- We expanded the plasma simulation code Cloudy to include a ~1000 reactions 68 species molecular network, including its effects on the ionization and thermal balance. Our treatment of sized-resolved grains and PAHs are described by van Hoof et al. (2004). These calculations can be applied to starbursts, ULIRGs, and O stars embedded in molecular clouds, as well as blister H II regions.

- We computed the ionization, thermal, and chemical structure of a large number of clouds, assuming constant gas pressure to relate the H II region and PDR. We present grids of results that show how emission lines vary over a wide range of $T_*$, $n_{H^+}$ and $U$.

- We predict the contribution of the H II region to PDR emission-line diagnostics. We find that the contribution from the ionized gas is more important for low-density, low-$G_0$ regions. The H II region contributes a larger percentage of the line emission when $T_*$ is high, since this increases the size of the H II region relative to the PDR.

- We use our calculations to determine the conditions in two star-forming regions: NGC 253 and NGC 7714. Analysis of the full spectrum, including lines from both the H II region and PDR, can self-consistently determine physical conditions.

- It will be possible to use H II region and PDR line ratios to determined densities in each region. When combined with either theoretical or observational measures of the gas temperature, the gas pressure can be obtained in each region. This can then test whether the cloud equation of state is dominated by gas pressure, as seems to be the case in the two objects analyzed here, or whether other pressure terms, perhaps magnetic pressure (Heiles & Crutcher 2005), dominate. The latter is frequently found across the galaxy.

Acknowledgements – N.P.A would like to acknowledge the Center for Computational Sciences at the University of Kentucky for computer time and financial support through a CCS fellowship. NSF and NASA supported this work through grants AST 03 07720, NAG5-12020, and HST-AR-10316.01-A. Conversations with Will Henney and Robin Williams helped guide our understanding of flows from molecular clouds. We would like to acknowledge the many useful discussions with the participants of the PDR workshop we attended in Leiden, The Netherlands. Finally, we would like to thank the anonymous referee for their careful reading of the manuscript.

# Appendix A

## Chemical Network in Cloudy—Reactions and Rate Coefficients

As shown in Figure 14, our calculations can extend to the point where the vast majority of elements are in molecular form. One of the major improvements to the spectral synthesis code Cloudy is upgrading the molecular network in Cloudy. This network consists of ~1000 reactions, which predict abundances for the following 68 molecules:

$H_2$, $H_2^*$, $H_2^+$, $H_3^+$, $HeH^+$, $CH$, $CH^+$, $OH$, $OH^+$, $O_2$, $CO$, $CO^+$, $H_2O$, $H_2O^+$, $O_2^+$, $H_3O^+$, $CH_2^+$, $CH_2$, $HCO^+$, $CH_3^+$, $SiH_2^+$, $SiH$, $HOSi^+$, $SiO$, $SiO^+$, $CH_3$, $CH_4$, $CH_4^+$, $CH_5^+$, $N_2$, $N_2^+$, $NO$, $NO^+$, $S_2$, $S_2^+$, $OCN$, $OCN^+$, $NH$, $NH^+$, $NH_2$, $NH_2^+$, $NH_3$, $NH_3^+$, $NH_4^+$, $CN$, $CN^+$, $HCN$, $HCN^+$, $HNO$, $HNO^+$, $HS$, $HS^+$, $CS$, $CS^+$, $NO_2$, $NO_2^+$, $NS$, $NS^+$, $SO$, $SO^+$, $SiN$, $SiN^+$, $N_2O$, $HCS^+$, $OCS$, $OCS^+$, $C_2$, and $C_2^+$.

We solve a system of linear, time steady equations for the molecular abundances (Atkins 1998). Consider a reaction of the form:

$$aA + bB \rightarrow cC + dD \tag{A1}$$

Here A, B, C, and D can be a molecule, atom, electron, or a photon and the lowercase letters are the stoichiometric coefficients. This reaction has the rate:

$$-kn(A)n(B) = \frac{1}{a}\frac{dn(A)}{dt} = \frac{1}{b}\frac{dn(B)}{dt} = -\frac{1}{c}\frac{dn(C)}{dt} = -\frac{1}{d}\frac{dn(D)}{dt} \quad (\text{cm}^{-3}\,\text{s}^{-1}) \tag{A2}$$

Many reactions will create or destroy a given species, and so the total rate of change of each species will be a sum over many reaction rates.

When the reaction rate contains the product of two densities, both of which are unknown, the equations are *non-linear*. We can linearize the rate by making the difference between the old and new solution small. We assure this by choosing the zone thickness $\delta r$ so that the conditions in adjacent zones never change by more than a set tolerance.

Using as an example the reaction rate from above, $-kn(A)n(B)$, we can define the difference between the old and new solution as:

$$\Delta_{A,B} = n(A,B)_{new} - n(A,B)_{old} \quad (\text{cm}^{-3}) \tag{A3}$$

The subscripts *new* and *old* refer to the current (unknown) solution and the previous (known) solution, respectively. In this notation, $n(A,B)_{new}$ is the density that goes into the reaction rate. Using A3, the reaction rate becomes:

$$-kn(A)n(B) = -k\left[\left(\Delta_A + n(A)\right)\left(\Delta_B + n(B)\right)\right] =$$
$$-k\left[\Delta_A\Delta_B + \Delta_A n(B)_{old} + \Delta_B n(A)_{old} + n(A)_{old} n(B)_{old}\right] = \quad \text{(cm}^3\text{ s}^{-1}\text{)} \quad \text{(A4)}.$$
$$-k\left[n(A)_{new} n(B)_{old} + n(A)_{old} n(B)_{new} - n(A)_{old} n(B)_{old}\right]$$

The last term was derived by ignoring terms to second order in delta. This expression for the rate coefficient now is linear, and allows for a solution through standard matrix inversion methods.

The chemical network is sensitive to details. We have included a complete treatment of the $H_2$ molecule in Cloudy (Shaw et al., 2005). This determines populations of 1893 levels producing 524,387 emission-lines, along with a detailed treatment of the self-shielding of $H_2$ electronic transitions. We also use other self-shielding functions. The calculations presented in this paper use the self-shielding function of Draine & Bertoldi (1996). Like $H_2$, CO is destroyed by absorption in electronic transitions, which we treat by using the dissociation rate given by Hollenbach, Takahashi, & Tielens (1991) with a rate coefficient derived by van Dishoeck and Black (1988). We determine the abundance of excited $H_2$ ($H_2^*$) following Tielens & Hollenbach (1985). In this approach, reactions form both $H_2$ and $H_2^*$, with the temperature barrier removed for reactions with $H_2^*$. Excited $H_2$ is very reactive, and, at shallow depths where the Solomon process produces a large population in $H_2^*$, its inclusion changes chemical equilibrium. In deeper regions, where $H_2$ is well shielded, nearly all $H_2$ is in the ground state.

We can include the effects of CO depletion on grain surfaces, using the treatment given in Hasegawa & Herbst (1993), but did not treat this process in our calculations.

We use the UMIST database (Le Teuff, Millar, & Markwick, 2000) for the vast majority of our rates. There are exceptions, however. For instance, an important reaction that forms CO involves $C^+$ and OH

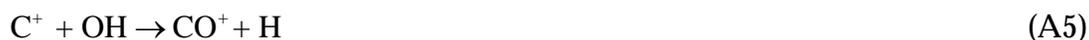
$C^+ + OH \rightarrow CO^+ + H$ (A5)

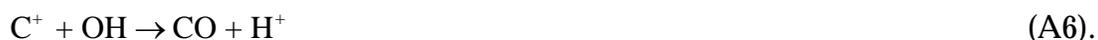
$C^+ + OH \rightarrow CO + H^+$ (A6).

In UMIST, these reactions have a rate, $k = 7.7 \times 10^{-10}$ cm$^{-3}$ s$^{-1}$, which does not depend on *T*. Table 1 of Dubernet et al. (1992) shows that, over their range of study, *T* < 100 K, the rate does depend strongly on *T*. We have derived a power law fit to the Dubernet et al. (1992) data, which fits their data for *T* < 100 K. Written in the same format as the UMIST, our derived rate is:

$$k = 2.7 \times 10^{-9} \left(\frac{T}{300}\right)^{-0.3508} \quad [\text{cm}^3 \text{ s}^{-1}] \tag{A7}$$

For the important charge transfer reactions between O and H, we use rates based on Stancil et al. (1999) that extends to $T = 10^7$ K. For $\text{H}^+ + \text{O} \rightarrow \text{H} + \text{O}^+$, the UMIST rate diverges from the rate derived by Stancil et al. for $T > 10,000$ K, while the reaction $\text{H} + \text{O}^+ \rightarrow \text{H}^+ + \text{O}$ diverges for $T > 41,000$ K. Cloudy is widely used to calculate conditions in hot plasmas, so it is necessary to have rates that extend over all physical regimes.

We compute photo-interaction rates by integrating the local continuum over the photo cross section, as

$$k = \int \frac{4\pi J_\nu}{h\nu} \sigma_\nu \, d\nu \quad [\text{s}^{-1}] \tag{A8}$$

Here the integral is over the relevant photon energies, $\sigma_\nu$ is the photo-cross section, and $J_\nu$ is the local radiation field (Figure 2). This radiation field includes the attenuated incident continuum and diffuse emission produced within the H II region and PDR. This allows such processes as photo-interactions with L$\alpha$ or the IR continuum produced by hot grains in the H II region to be self-consistently included.

Two important rates, H radiative recombination and He charge transfer recombination with H, are significantly different from the UMIST database. The UMIST rate of reaction $R^H$ for $\text{H}^+ + \text{e} \rightarrow \text{H} + h\nu$ is:

$$R^H_{rad-rec} = 3.50 \times 10^{-12} \left(\frac{T}{300}\right)^{-0.75} \quad 10 < T < 20,000 \quad [\text{cm}^3 \text{ s}^{-1}] \tag{A9}$$

This differs from the exact rates computed by Ferland et al. (1992) by ~30% for $T < 100$ K. This is important because, even though H$^+$ is only a trace species in a PDR, its absolute abundance is large. It can be a significant electron donor, and this affects molecular reaction rates and therefore molecular formation. An approximation to the radiative recombination rate, over the UMIST $T$ range, is:

$$R^H_{rad-rec} = 3.50 \times 10^{-12} \left(\frac{T}{300}\right)^{-0.75} - 3.0 \times 10^{-12} \left(\frac{T^{1.05}}{40}\right)^{-1.3} \quad [\text{cm}^3 \text{ s}^{-1}] \tag{A10}$$

The first term is just the UMIST rate, while the second term is a correction factor that allows for better agreement at low $T$. Figure 34 shows how the Cloudy recombination rate, UMIST rate, and the best fit to the Cloudy recombination rate vary over the range 10-20,000 K. The rate becomes negative for $T < 2$ K. The lower bound of $T$ for most UMIST rates is 10 K, so our rate is stable over the range covered by UMIST.

The UMIST rate for the charge transfer process $He^+ + H \rightarrow H^+ + He$ is a factor of 4 too large. This rate in UMIST should be:

$$R_{ch-tr}^{He} = 1.21 \times 10^{-15} \left(\frac{T}{300}\right)^{0.18} \quad 10 < T < 1000 \quad [\text{cm}^3 \text{ s}^{-1}] \tag{A11}.$$

The original rate was derived by Zygelman et al. (1989), while Stancil et al. (1998) give the correct rate. Charge transfer recombination of He with H is the main formation process for $He^0$ at the illuminated face of a PDR. The large rate leads to a factor of four decrease in the density of $He^+$, affecting the abundances of other molecules due to reactions of $He^+$. This primarily affects molecular abundances for low $A_V$, or translucent clouds. For high $A_V$, the abundance of $H^0$ is significantly less. This makes this rate less important at high $A_V$.

# Appendix B
## Charge and Energy Exchange of Gas and Dust

When an atom or ion collides with a grain particle, there is a possibility that an exchange of electrons occurs. We will assume that the atom or ion has charge $Z$ before it hits the grain, and that after the charge exchange it leaves the grain with charge $Z_0$. The value of $Z_0$ is determined by assuming that electrons always move into the deepest potential well (i.e., the exchange is always exothermic). This implies that electrons can move in either direction, although in most cases they will move from the grain into the ion. The exchange of electrons will continue until the next exchange would have been endothermic. The ion charge where this condition is reached is by definition $Z_0$. This definition is deceptively simple as the depth of the potential well of the grain and the ion must be reevaluated each time an electron is exchanged. Expressions for the grain potential were taken from Weingartner & Drain (2001) and will be discussed in more detail in a forthcoming paper. In most cases $Z_0 = 0$, but for third and fourth row elements with low ionization potentials there may be exceptions. Note that $Z_0$ depends on $A$ and $Z$, as well as the change of grain $Z_i$. The rate (per unit projected grain surface area) for this process is given by

$$\alpha_p(A, Z \to Z_0) = N_{A,Z} \overline{v_A} S_p \sum_{i=1}^{n} f_i \eta_Z, \text{ with } Z \geq 0, \text{ [cm}^{-2}\text{ s}^{-1}\text{]} \qquad (B1)$$

where $f_i$ is the fraction of the grains that have charge $Z_i$ (see van Hoof et al. 2004 for a detailed discussion of the way the grain charge distribution is derived). $N_{A,Z}$ is the number density of the incoming ion, $A^{Z+}$, $\overline{v_A}$ is the average velocity of that ion, and $S_p$ is the sticking probability for atoms and ions (assumed to be 1 when $Z \neq Z_0$). When an ion approaches a charged grain, there will be either attraction or repulsion depending on the sign of the charges. This Coulomb interaction will either magnify or diminish the collisional cross section, an effect which is described by the factor $\eta_Z \equiv \tilde{J}(\tau, \nu)$. We use the expressions given in Draine & Sutin (1987), which include polarization effects in the grain induced by the approaching particle. This theory needs two scaling parameters $\tau$ and $\nu$ which are simply the ratio of the two charges ($\nu$) and of the kinetic energy of the approaching particle and the potential energy of the grain ($\tau$). These parameters are given by

$$\tau = \frac{4\pi\varepsilon_0 a k T_e}{(Ze)^2} \qquad (B2)$$

$$\nu = \frac{Z_i}{Z} \tag{B3}$$

where *a* is the radius of the grains, and the rest of the symbols have their usual meaning.

Charge exchange between grains and negatively charged ions as well as charged molecules is currently not implemented. These will be added at a later date.

The Cloudy treatment of collisional heating or cooling of the gas due to collisions with the grains was first described in Baldwin et al. (1991). The formulation given in that paper has been largely retained, with the following two exceptions. It is now assumed that the outgoing particle has charge $Z_0$ instead of always being neutral, as was outlined above. Secondly, the Coulomb interaction between the ion and the grain is treated using expressions from Draine & Sutin (1987). The factor $\eta_Z$ needed in that formulation has already been discussed, the factor $\xi_Z$ has been replaced by $\tilde{\Lambda}(\tau,\nu)/2$. Since the heating rates of the gas depend on the grain charge, all changes to the charging physics will affect the heating rates as well. These changes have been partially discussed by van Hoof et al. (2004) and largely follow Weingartner & Draine (2001).

The most important changes with respect to Baldwin et al. (1991) are the introduction of the hybrid grain charge model (which includes a minimum charge for all grains and introduces a bandgap between the valence and conduction band for silicates), the reduction of the work function of graphite, an upgrade of the treatment of the photoelectric effect, and an altered prescription for the sticking efficiency of electrons. A comprehensive discussion of the grain treatment currently included in Cloudy will be presented in a forthcoming paper.

# Appendix C

## Cosmic Ray Heating

Cosmic rays are the primary source of heating and ionization deep in molecular clouds. Cosmic rays of primary energy $E_0$ ionize atoms and molecules, primarily hydrogen and helium. The ejected secondary electrons can produce further ionization. The heating and ionization rates depend on the primary ionization rate, electron fraction, and energy of the secondary electron.

Calculations by Shull & van Steenberg (1985, Fig. 3) describe the number of secondary ionizations per primary ionization and the heating efficiency due to cosmic rays. Wolfire et al. (1995) use an equation from Binette, Dopita & Tuohy (1985) that is based on this work, but modified to extend to lower $E_0$. We also use these results, with the primary cosmic ray energy $E_0$=35 eV as in Wolfire et al (1995). The form of their equation for the heating efficiency, $\varepsilon(E,x)$, is:

$$\varepsilon(E,x) = \beta_1(x) + \left[\frac{g(E)}{\alpha_1(x)} + \frac{1}{1-\beta_1(x)}\right]^{-1} \tag{C1}$$

where $E$ is the primary cosmic ray energy, $x$ is the electron fraction, and $\beta_1(x), \alpha_1(x)$, and $g(E)$ are coefficients that depend upon $E$ and $x$.

We used Figure 3 of Shull & van Steenberg and derived a best-fit formula for $\varepsilon(E,x)$ as a function of $x$.

$$\varepsilon(35eV, x) = -8.189 - 18.394x - 6.608x^2 ln(x) + 8.322e^x + 4.961\sqrt{x} \tag{C2}$$

which is valid for electron fractions of $1 > x > 10^{-4}$.

Figure 35 compares C2 to the expression given by equation A3 of Wolfire et al. (1995). They agree to within 10%.

Cosmic rays effect physical conditions in ways that cannot be treated by only using the UMIST database. We include multi-level atoms of H (Ferguson et al. 1997) and He (Porter et al. 2005), including collisional radiative transfer processes. Cosmic rays excite resonance lines of both $H^0$ and $He^0$, and these are degraded into L$\alpha$ after multiple scatterings. The code computes the mean intensity in all lines, and this is included in the calculation of the radiation field in equation A5. This contributes to grain heating and to photoionization of $H^0$ and $He^0$, from the metastable $2s$ and $2\,^3S$ levels respectively. This increases the electron fraction.

# 9 Figures

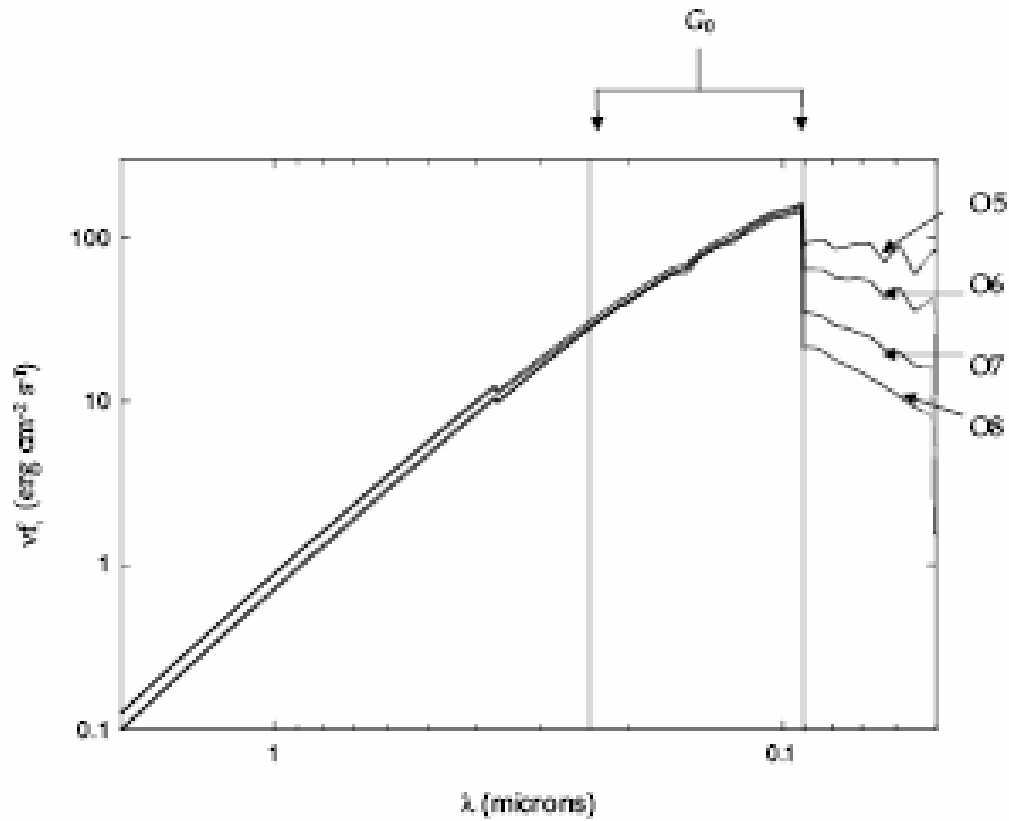

Figure 1 Dependence of the shape continuum on the stellar spectral classes considered in our calculations. The area between the vertical lines represents the continuum used to define $G_0$. The Lyman continuum depends strongly on spectral type, while $G_0$ is relatively constant. This causes the size of the H II Region relative to the PDR to vary, which affects the contribution of the H II region to PDR emission-line diagnostics.

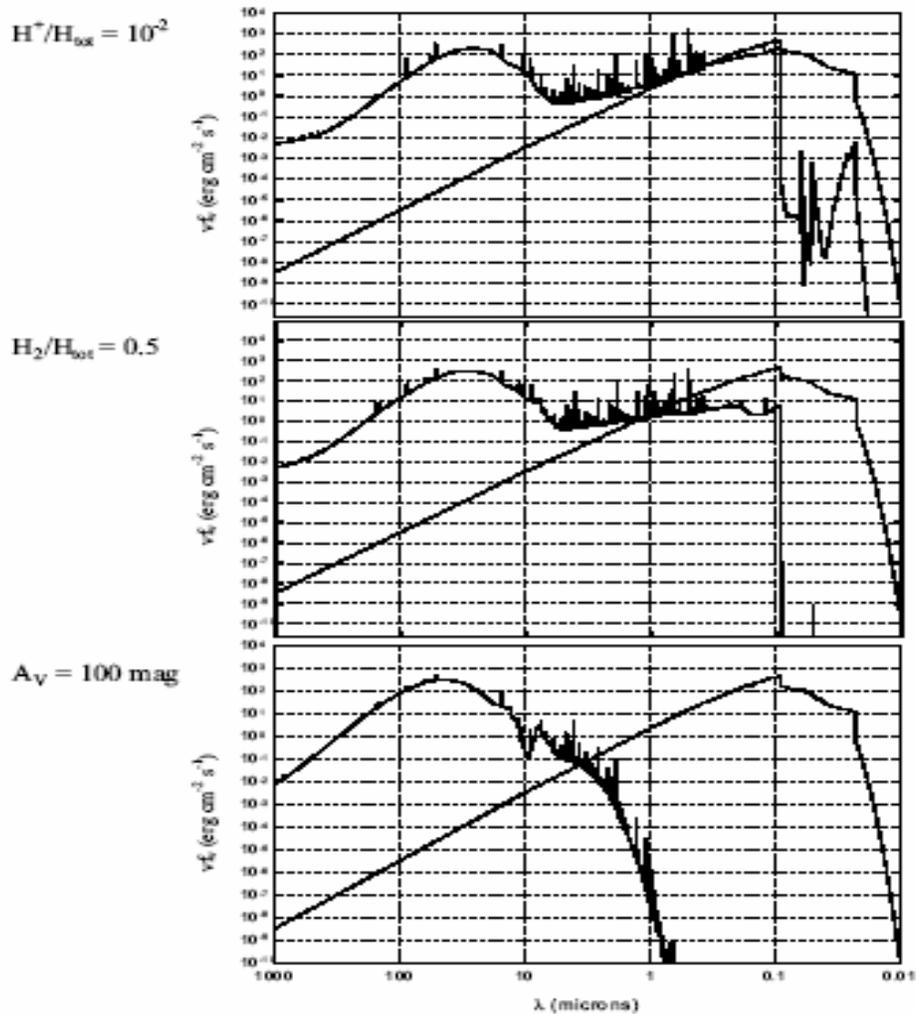

Figure 2  Continuum at three positions along a ray into a standard cloud.  The incident and net local continuum is shown at three positions; the hydrogen ionization front, the point where half the hydrogen is $H_2$, and the shielded face.  At the hydrogen ionization front the Lyman continuum is strongly absorbed by hydrogen and dust and reemitted in the infrared by dust.  PAH emission is present at the half-$H_2$ point, along with graphite absorption at 2175Å.  All radiation with $\lambda < 1\mu m$ is absorbed at the shielded face, the silicate feature is clearly seen in absorption, and a rich IR emission spectrum produced by atoms and molecules is observed.

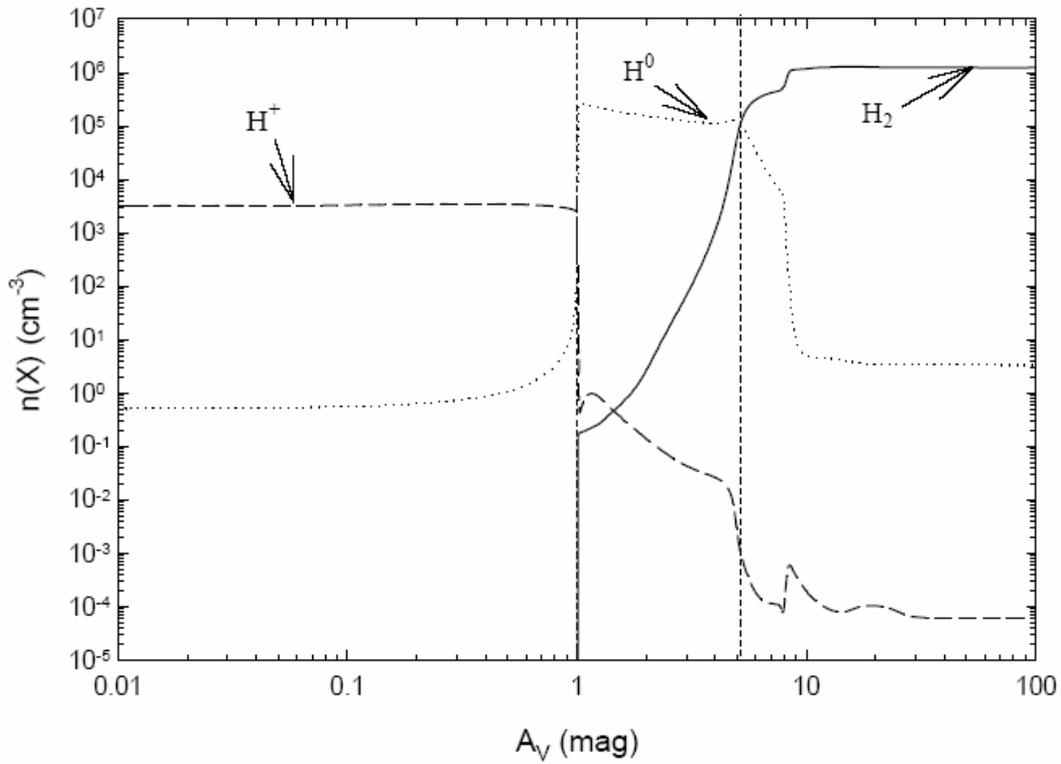

Figure 3 The $H^+$, $H^0$, and $H_2$ density structure of the standard cloud. This shows the classic transition from ionized to atomic and finally to molecular gas. From the illuminated face to the shielded face, $n_H$ increases by ~$10^3$ due to the assumption of constant gas pressure. The two vertical lines represent the hydrogen ionization front and the half-$H_2$ point.

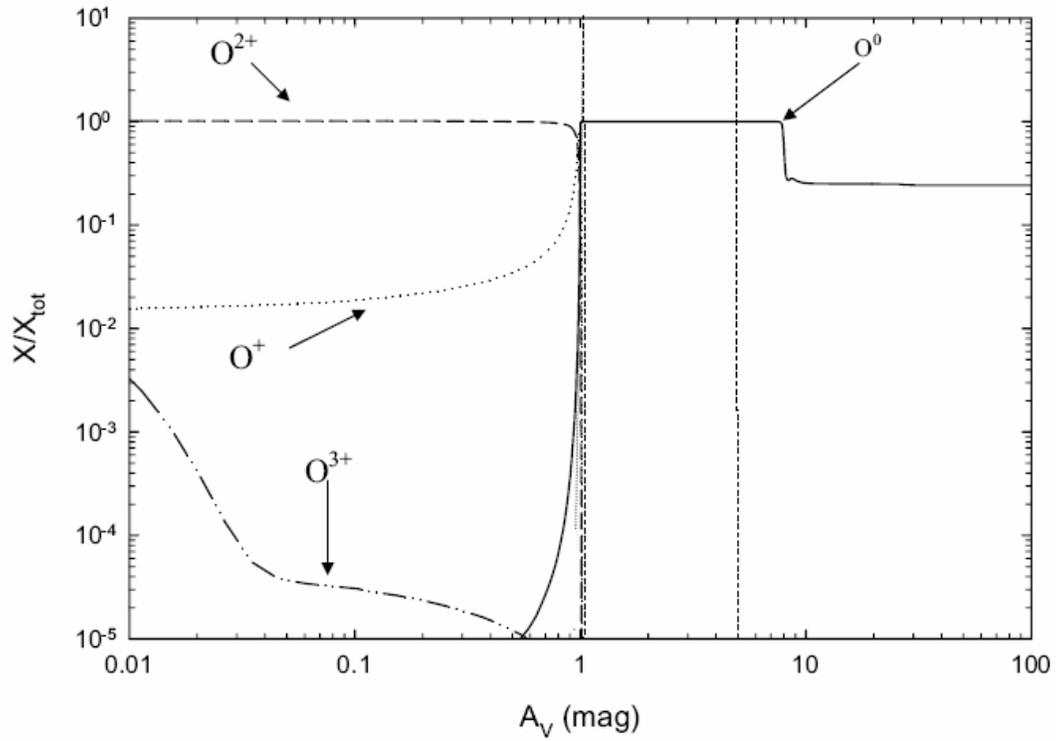

Figure 4 The oxygen ionization structure for our standard cloud. At the illuminated face, O is in the form of $O^{2+}$ and $O^+$. At the hydrogen ionization front, most O is in the form of $O^0$ due to rapid charge exchanges with $H^0$. the relative abundance of $O^0$ decreases deep in the cloud due to the formation of CO.

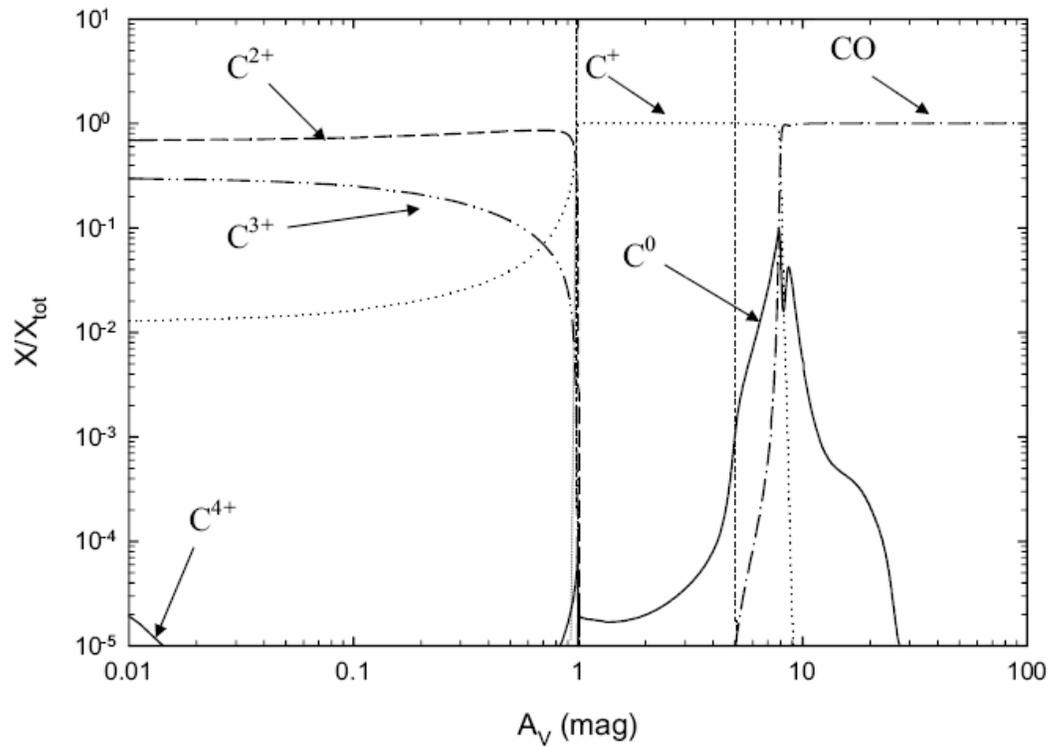

Figure 5 The carbon ionization structure for our standard cloud. At the illuminated face, C is predominately in the form of $C^{2+}$ and $C^{3+}$, with some $C^+$. Near the hydrogen ionization front carbon becomes predominately $C^+$. Once $H_2$ forms, carbon makes the classic PDR transition from $C^+$ to $C^0$ and quickly over to CO.

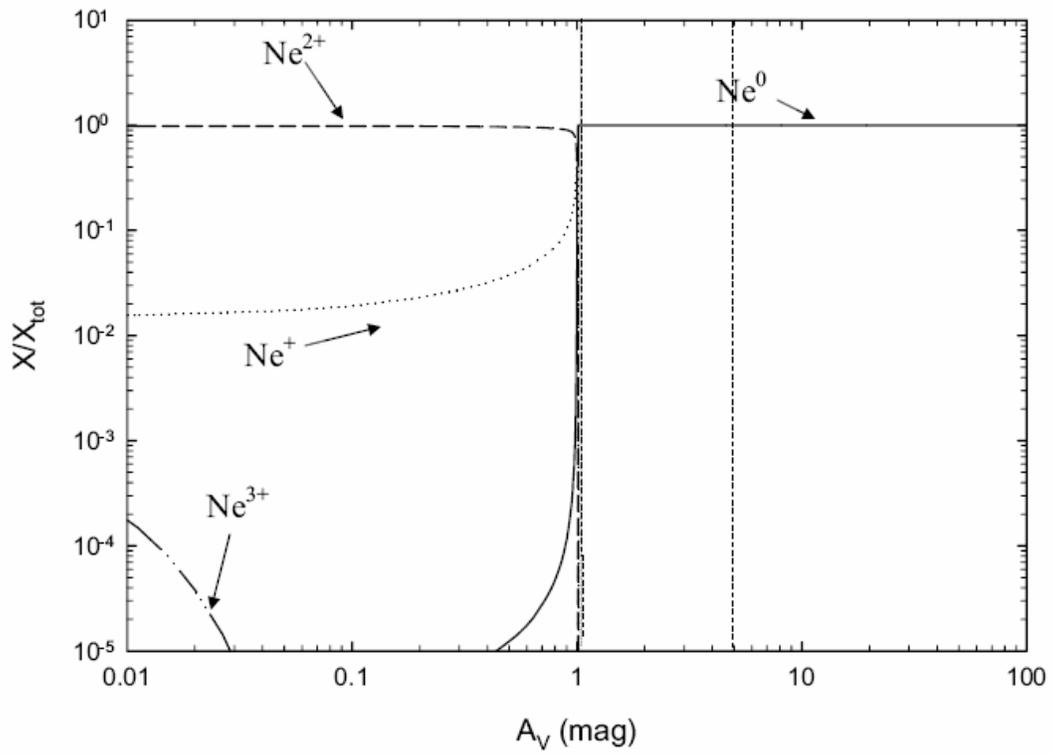

Figure 6  The neon ionization structure for our standard cloud.  Neon is singly and doubly ionized in the H II region, while it becomes atomic at the hydrogen ionization front.  $Ne^0$ has a closed shell, and therefore is not chemically active in the molecular cloud.

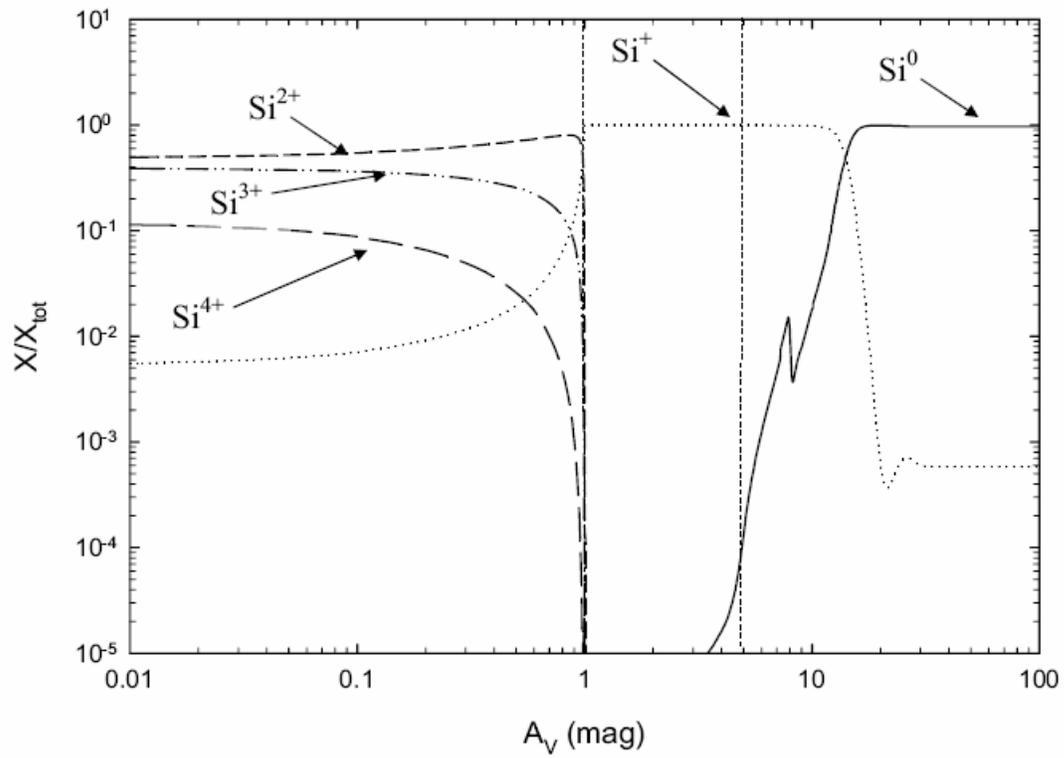

Figure 7 The silicon ionization structure for our standard cloud. Silicon is in the form $Si^{2+}$, $Si^{3+}$, and $Si^{4+}$ in the H II region As the hydrogen ionization front is approached, $Si^+$ becomes the dominant stage of ionization. $Si^+$ remains the dominant ionization stage through the PDR, becoming neutral at an $A_V$ of ~15.

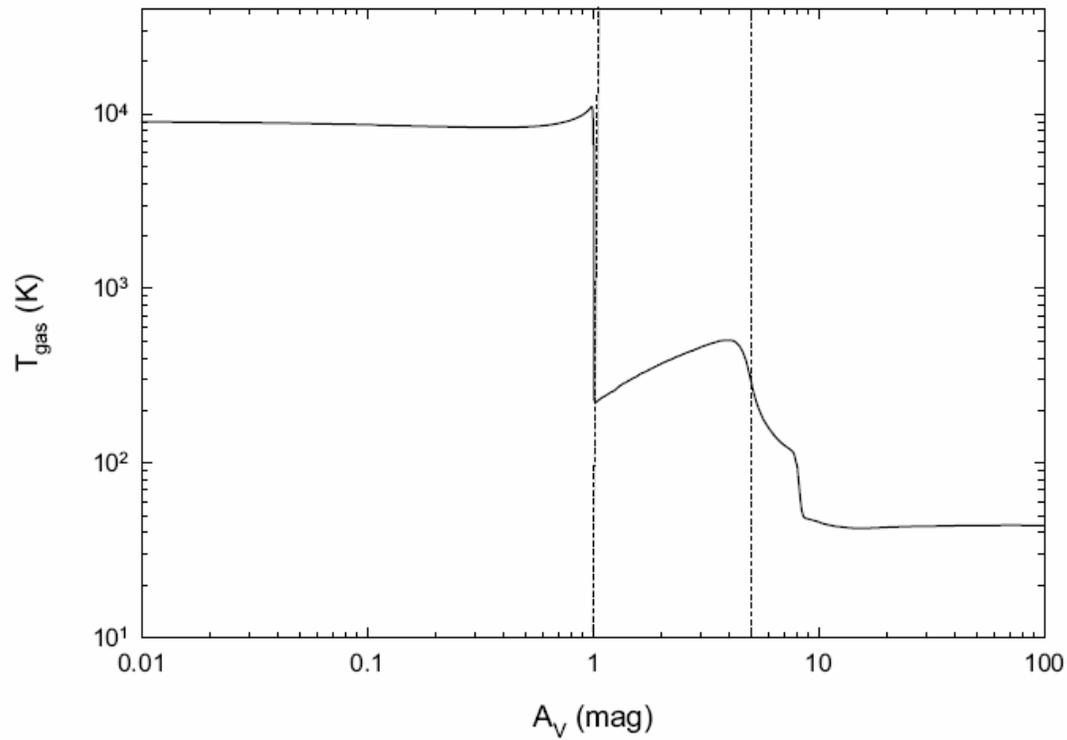

Figure 8 Gas temperature for our standard cloud. The temperature is ~$10^4$ K in the H II region,. At the ionization front, $H^0$ photoionization no longer heats the gas, and the temperature falls to a few hundred Kelvin. The temperature continues to fall as the gas becomes more molecular, eventually reaching a temperature of ~40 K at the shielded face.

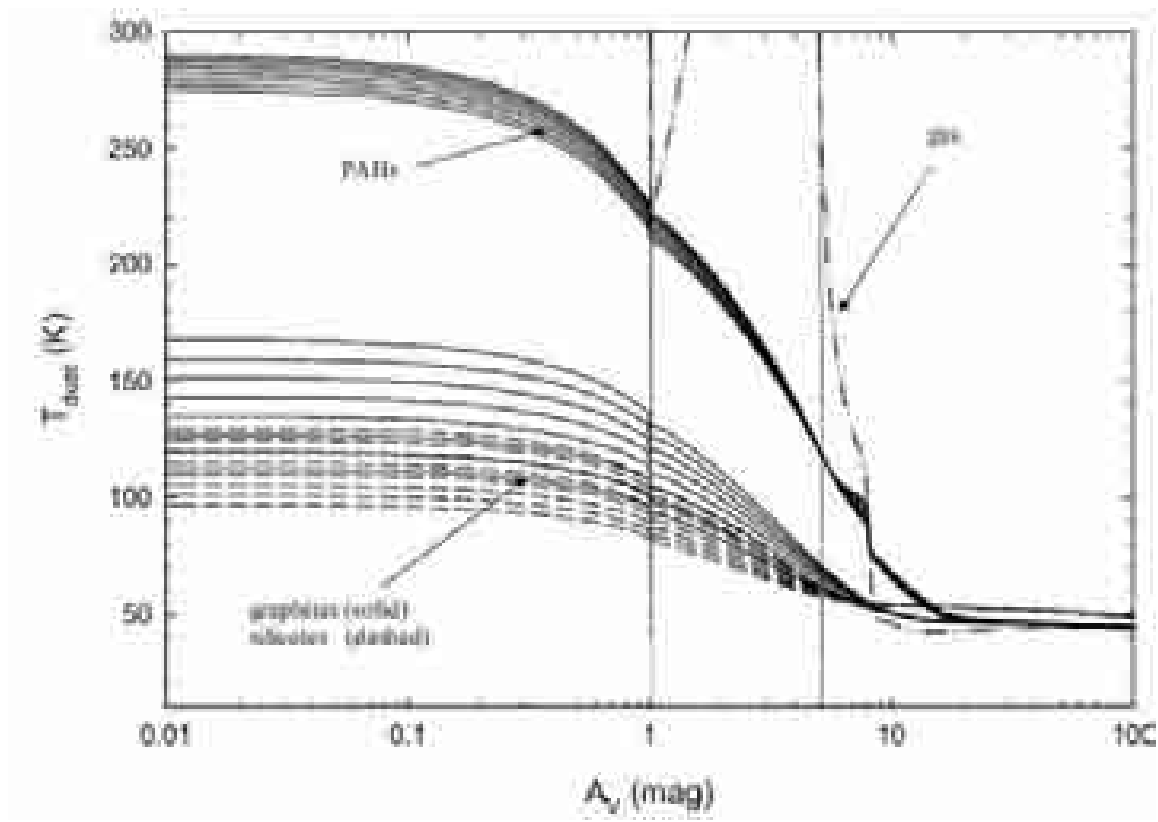

Figure 9  Dust temperature for all species and sizes considered in our standard cloud.  PAH temperatures are shown.  We assume that the PAH abundance is proportional to the atomic hydrogen fraction.  PAHs are the hottest grain species in the H II region but their abundance is negligible.  Additionally, graphite is somewhat hotter than the silicates in the H II region and PDR due to their optical properties.  Deep in the cloud, the gas and dust temperatures equilibrate, with the silicate dust temperature ~5 K higher (see Figure 15).

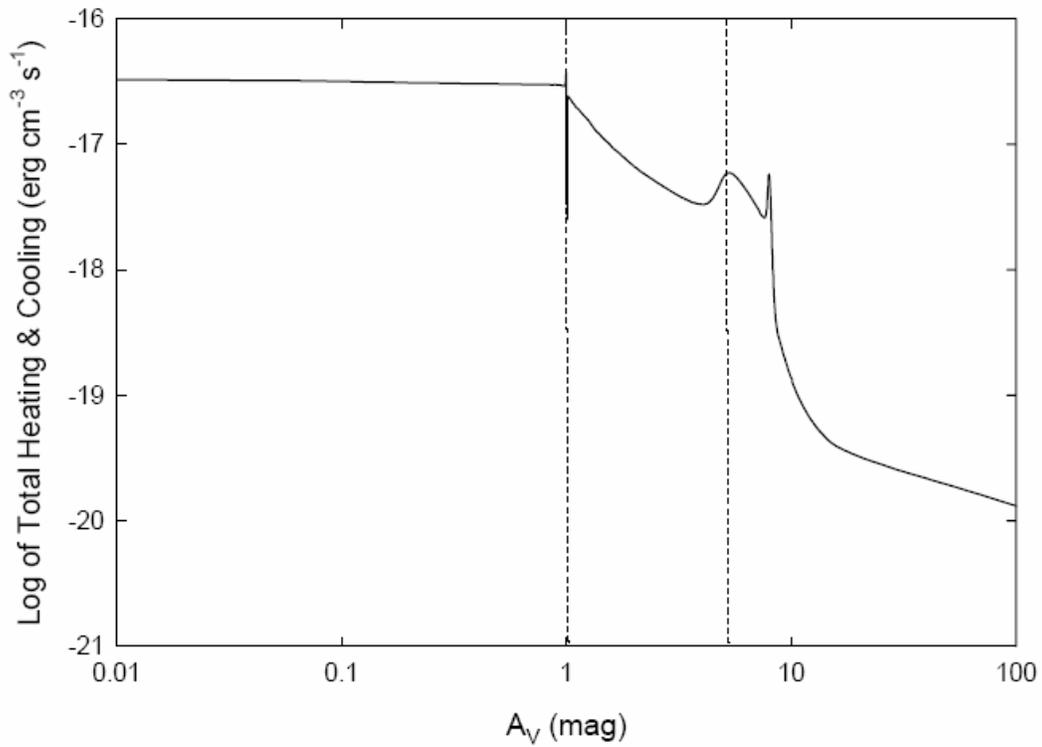

Figure 10 Total heating and cooling rate in our standard cloud. We assume energy balance so the two curves overlap. This plot can be compared with Figures 11 and 12 to determine the total heating or cooling rate for a given species.

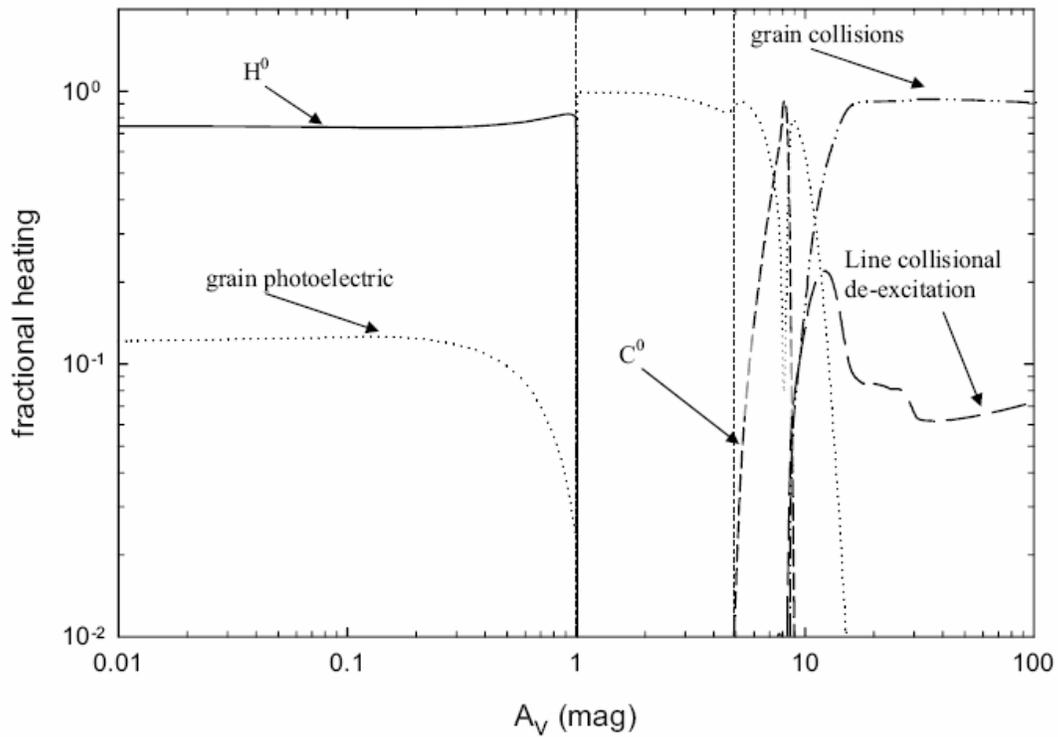

Figure 11  Important heating processes in our standard cloud.  In the H II region, heating is dominated by photoionization of $H^0$.  In the PDR, grain and carbon photoionization dominates.  Eventually, carbon and dust photoionizing photons are extinguished.  Deep in the cloud grain collisional heating (see Figure 15) and heating due to line collisional de-excitation are the primary heating agents.

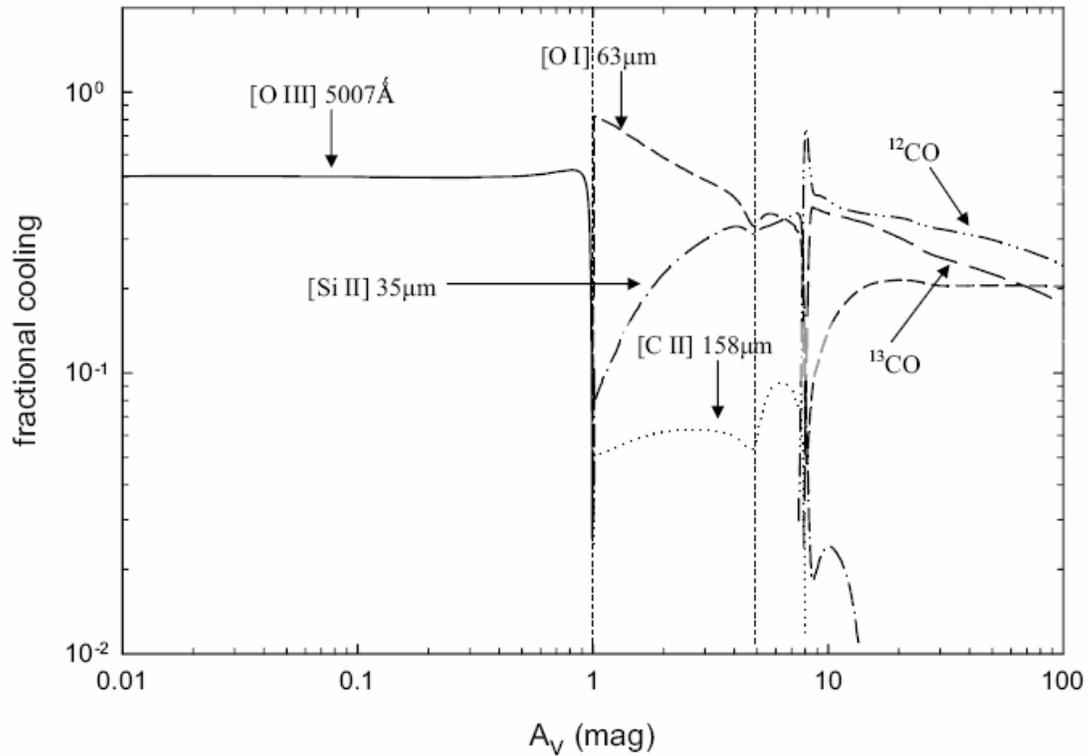

Figure 12 Important cooling processes in our standard cloud. Cooling is dominated by fine structure excitation followed by radiative de-excitation. The [Si II] 35 μm and [C II] 158 μm line peaks as hydrogen makes the transition from atomic to molecular. Deep in the calculation, CO is the primary coolant. Carbon monoxide cooling peaks in regions where carbon forms CO, and falls off deep in the cloud where all the CO is in the ground state.

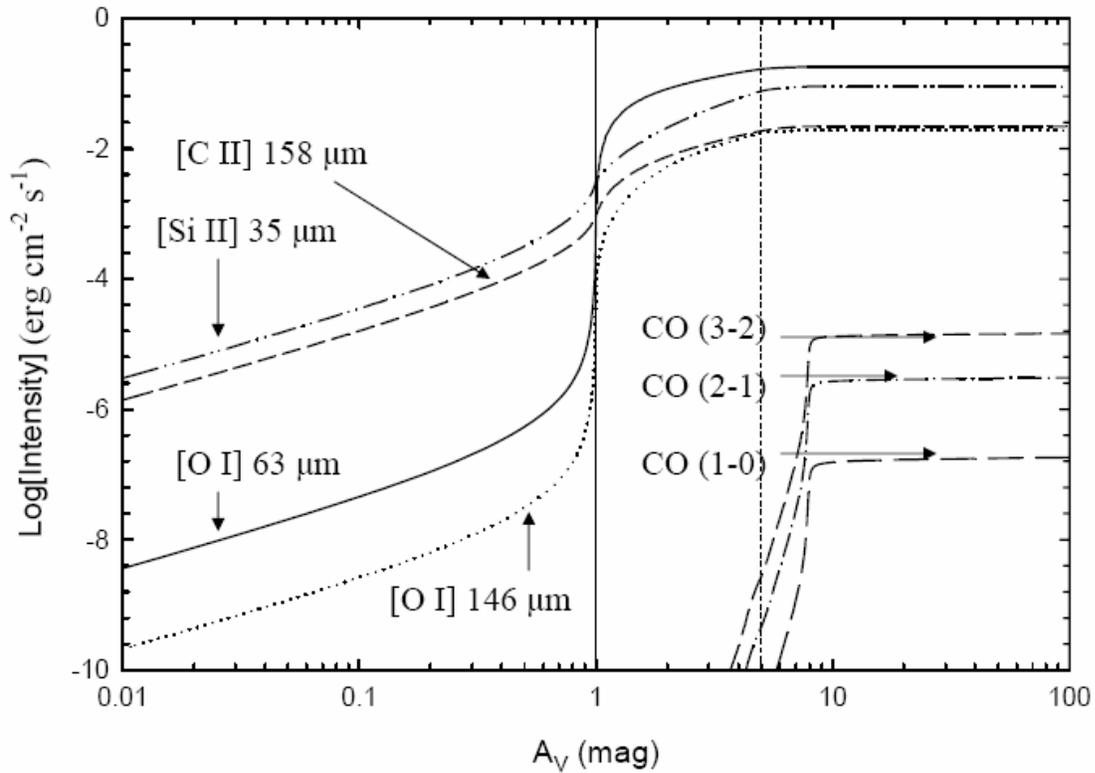

Figure 13 Integrated emission of PDR lines in our standard cloud. For our choice of $n_H$, $U$, and $T^*$, the assumption that low-ionization potential lines form in the PDR is valid. In the PDR. $C^+$, $Si^+$, and $O^0$ become the dominant stage of ionization, and collisions with $H^0$ excite their fine structure levels. CO emission from low J levels occurs at the $C^+/C^0/CO$ transition. Due to the high column densities of the low CO rotational levels, these lines quickly become optically thick

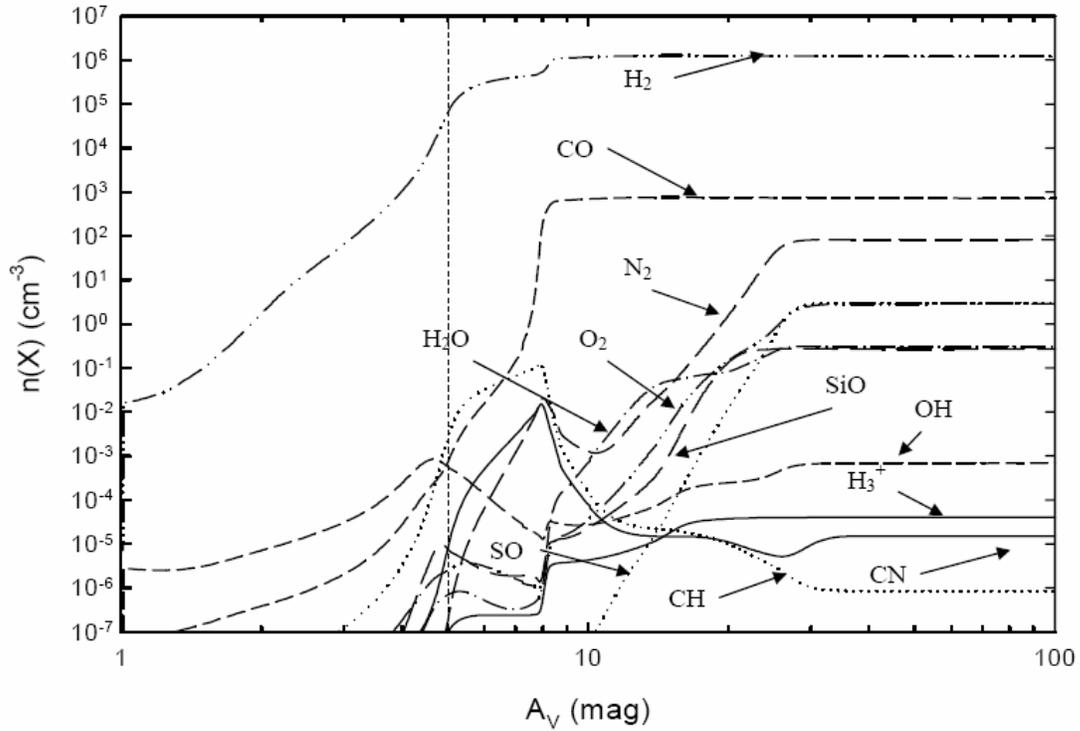

Figure 14  Molecular abundances in our standard cloud.  The formation of $H_2$ acts as a catalyst for other molecules, especially CO.  Once CO fully forms, other species start to form molecules.  At the shielded face, all the hydrogen is tied up in $H_2$, all carbon is in CO, all nitrogen is in $N_2$, and a large amount of oxygen is in CO, $O_2$, and $H_2O$.

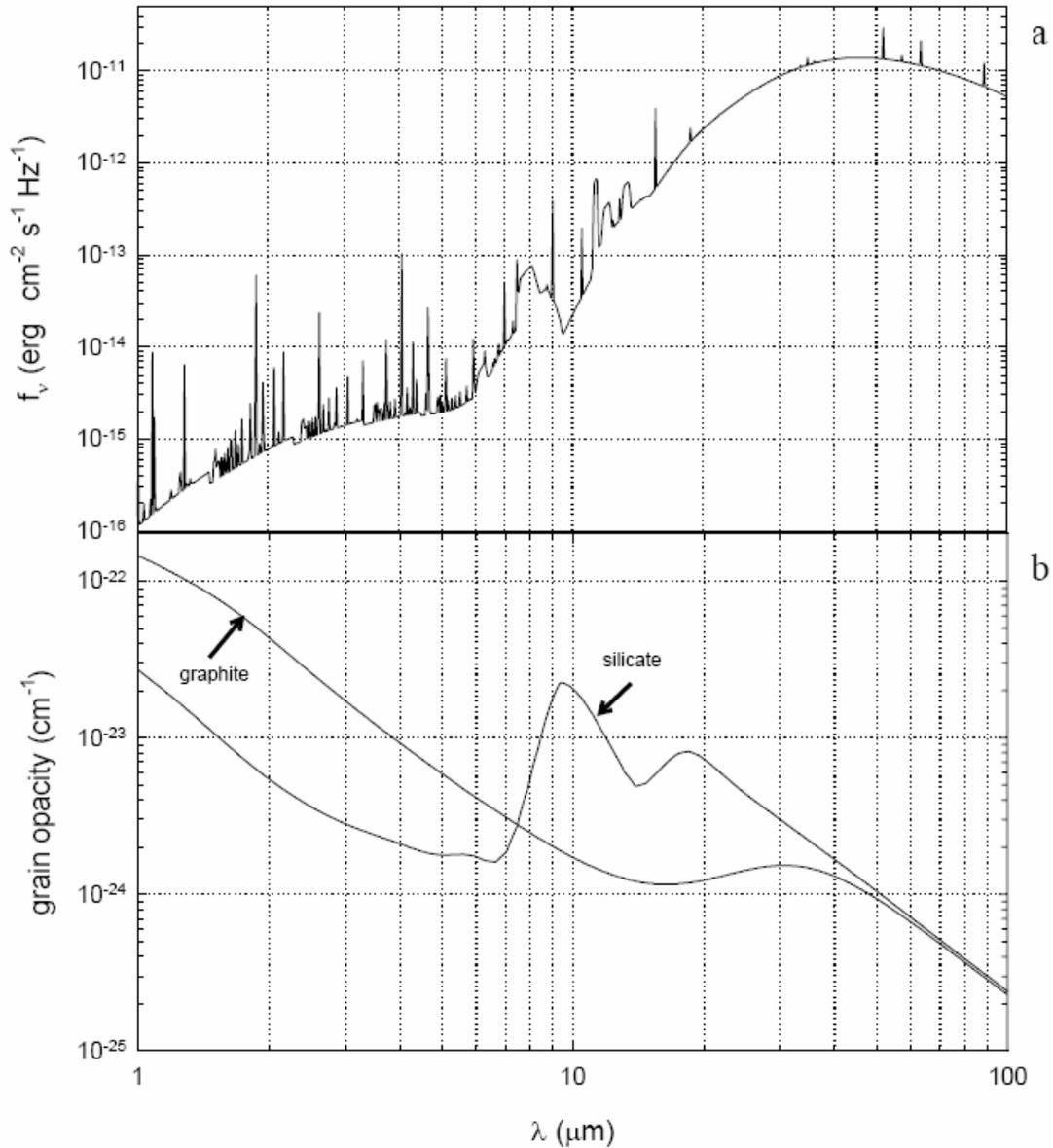

Figure 15 Local continuum and grain opacity at $A_v$ =20 in our standard cloud. This plot shows why the silicate dust is hotter than the graphite deep in the molecular cloud. The product of flux and opacity (integrated over frequency) is the heating rate. The large silicate opacity in regions of high infrared flux keeps the silicate grains ~5 K hotter than the gas or graphite deep in our calculation.

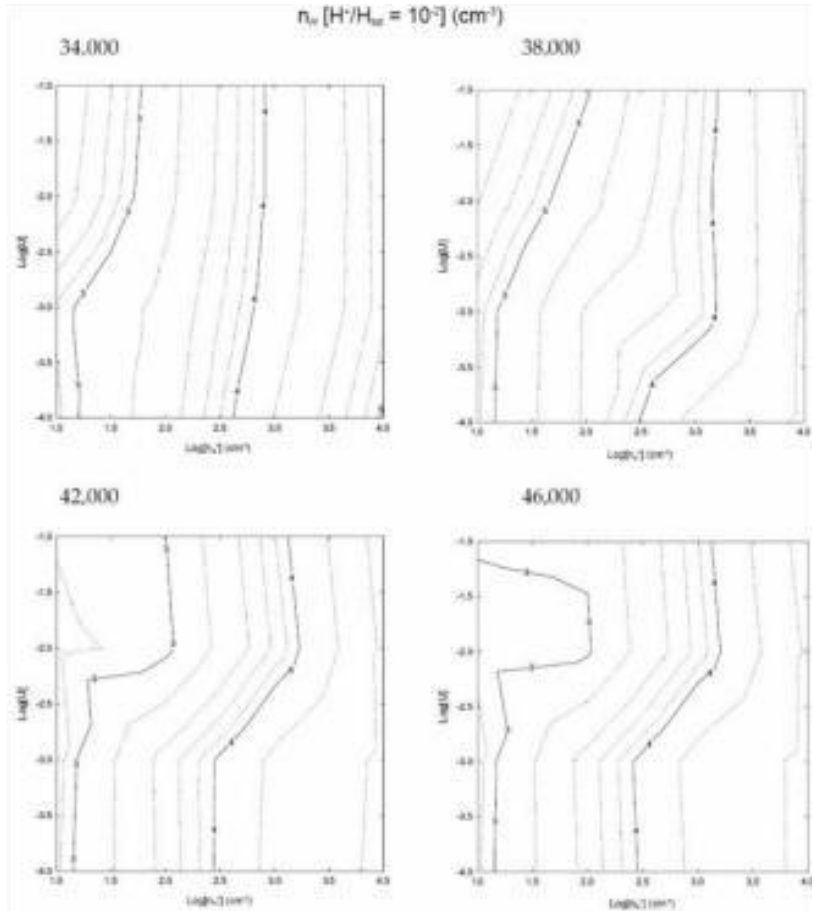

Figure 16 Log[$n_H$] at the point in our calculations where the $H^+/H_{tot}$ ratio falls to 1%. This is the hydrogen density at the face of the PDR. For a given $U$ and $n_{H^+}$ ($n_H = n_{H^+}$ at the illuminated face), For a given set of parameters the hydrogen density at the face of the PDR is roughly 1-2 orders of magnitude greater than at the H II region illuminated face. The temperature at the face of the PDR is typically $10^2$-$10^3$ K

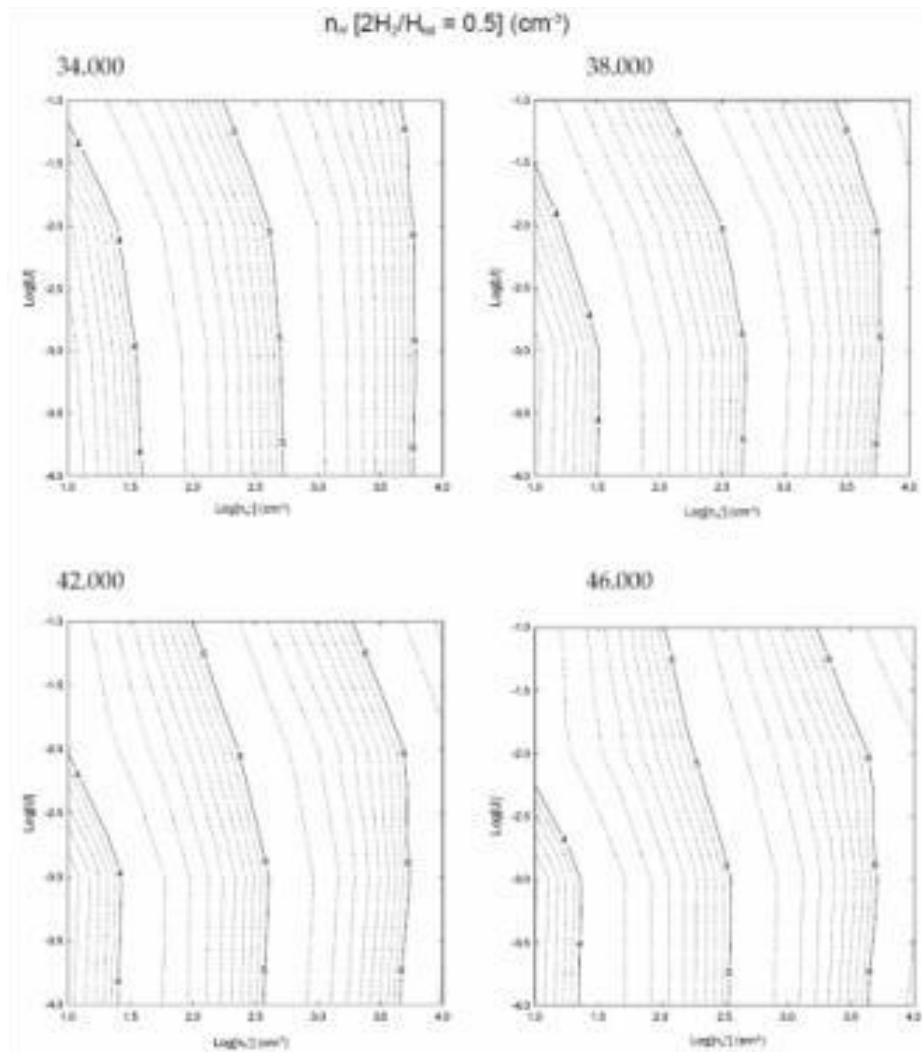

Figure 17 Log[$n_H$] at the point when $2H_2/H_{tot}$ reaches 0.9. Our constant-pressure calculations do not produce a single density in the PDR, but a typical PDR density would be an average of the density determined from this plot and Figure 16. The density at the $2H_2/H_{tot}$ = 0.9 point is ~2.5 orders of magnitude greater than the density at the illuminated face, and has a temperature of ~40 K.

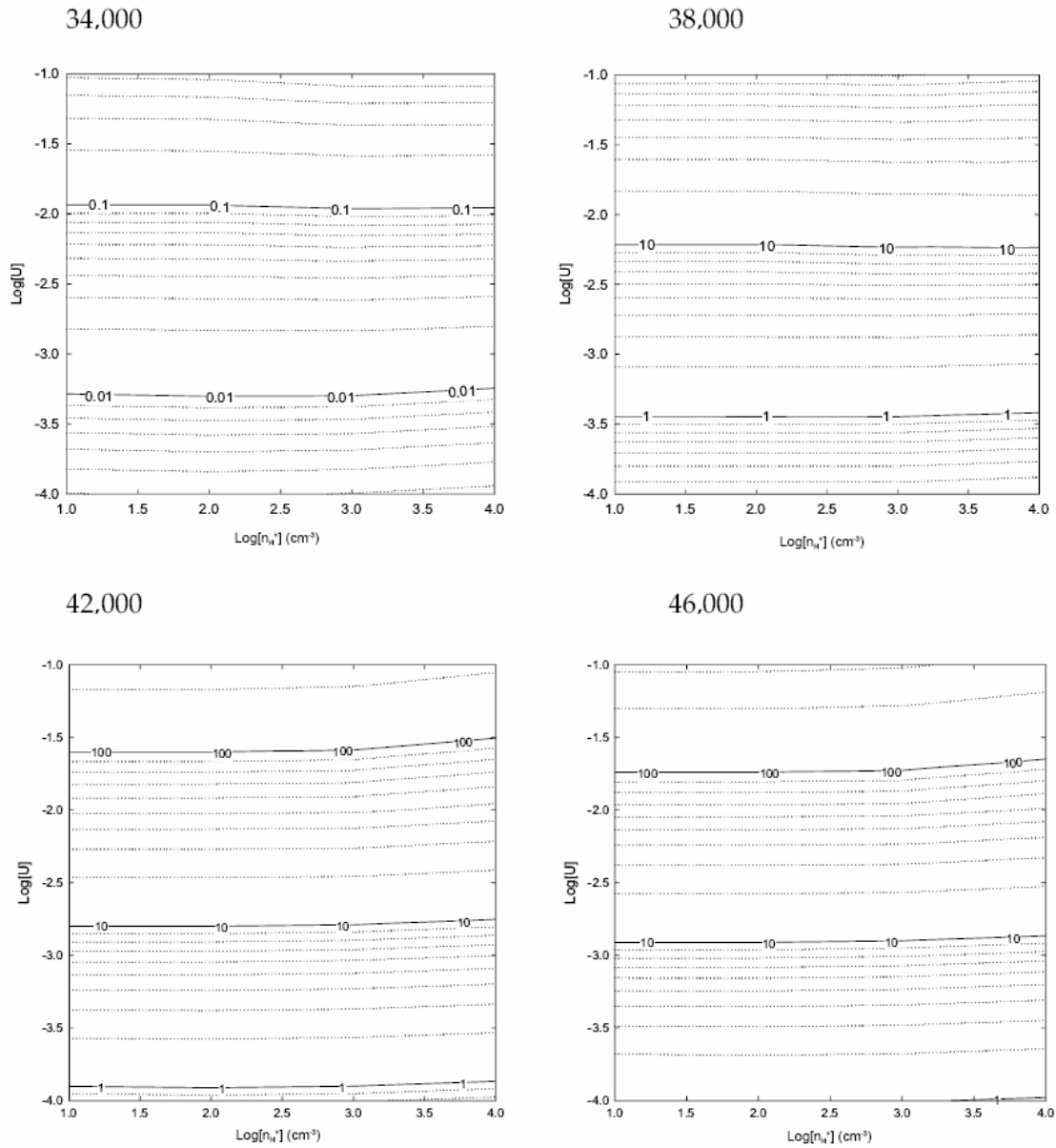

Figure 18 Ratio of the [Ne II] 15.55 μm line intensity to the [Ne III] 12.8μm intensity. Because of the large critical densities of these two lines (>$10^5$ cm$^{-3}$; Giveon et al 2002), this ratio is sensitive to $T^*$ and $U$ but not $n_{H^+}$ for a range of H II region densities. Both $T^*$ and $U$ can be determined by combining this with Figures 19-21,.

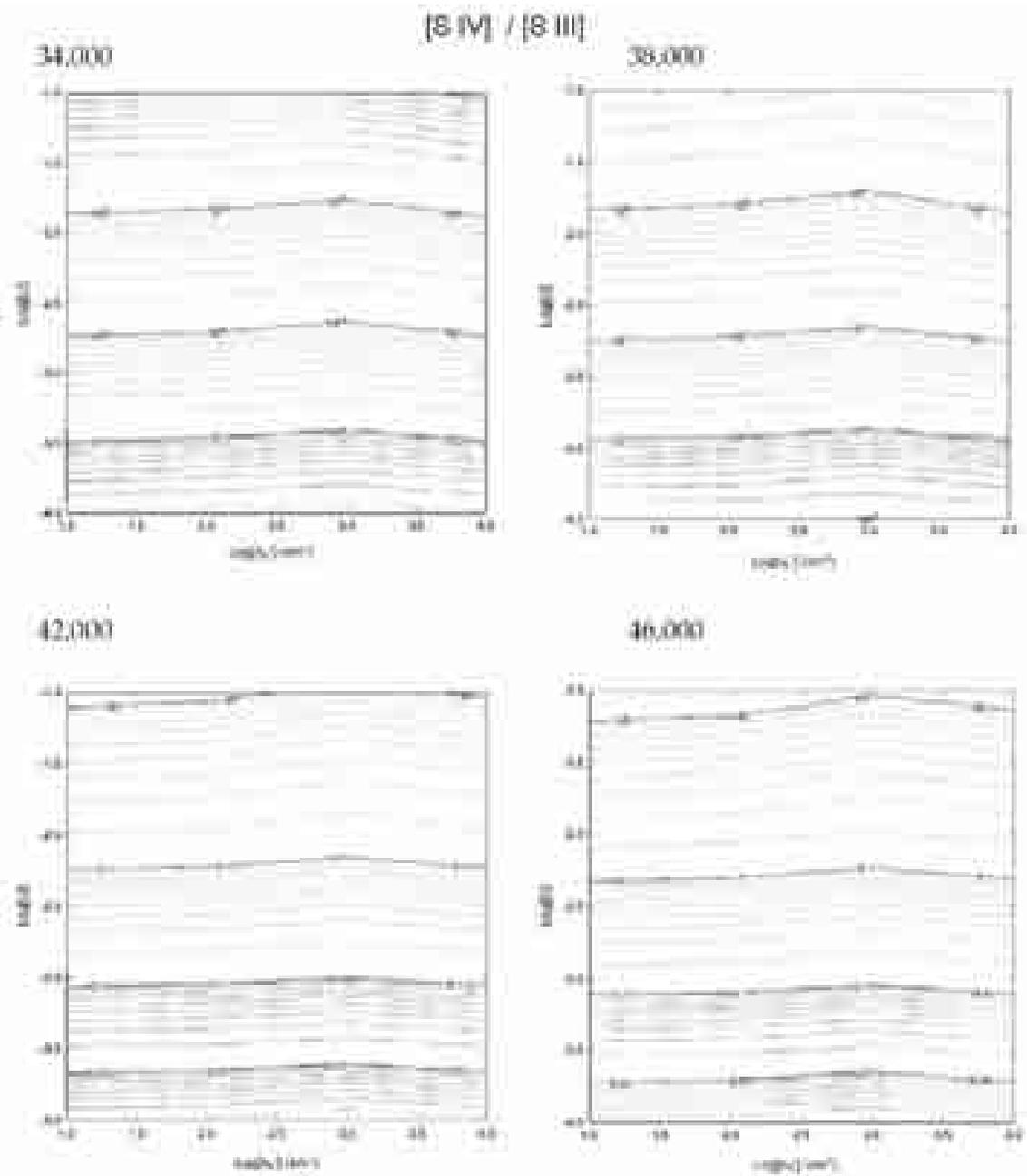

Figure 19 Ratio of the [S IV] 10.5 μm line intensity to the [S III] 18.7 μm intensity. The critical density of both of these transitions exceeds $10^4$ cm$^{-3}$ (Giveon et al. 2002), making the ratio of these lines insensitive to $n_{H^+}$ in our calculations. The best way to determine $T$ and $U$ would be to combine calculations such as those in Figures 18 - 21 with observations (see Giveon et al. 2002, Morisset 2004).

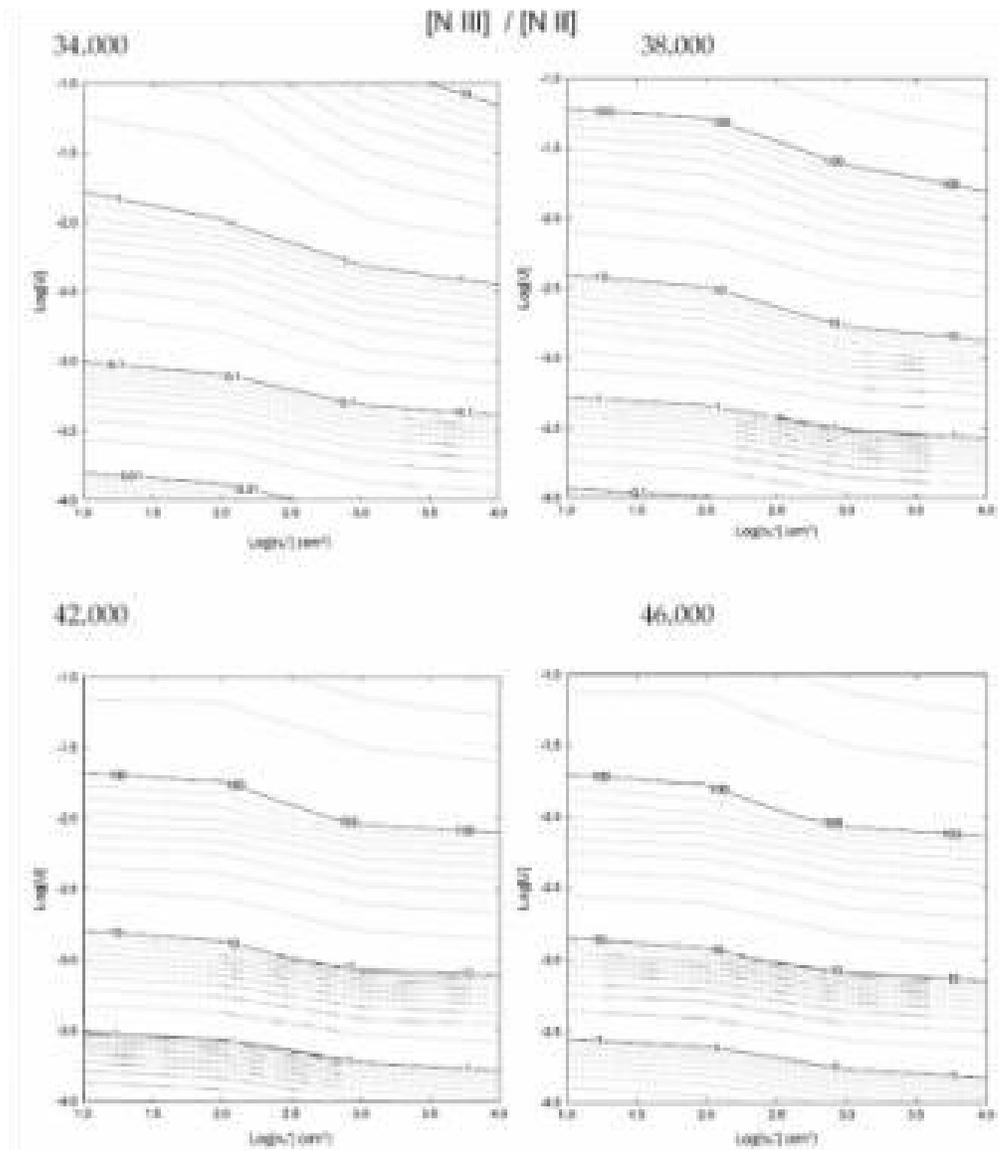

Figure 20 Ratio of the [N III] 57 μm intensity to the [N II] 122 μm intensity. The critical densities for these transitions are less than $10^4$ cm$^{-3}$ (Malhorta et al. 2001). This means that, for the range of $n_{H^+}$ considered in our calculations, this ratio can also depend on density in addition to $T$ and $U$.

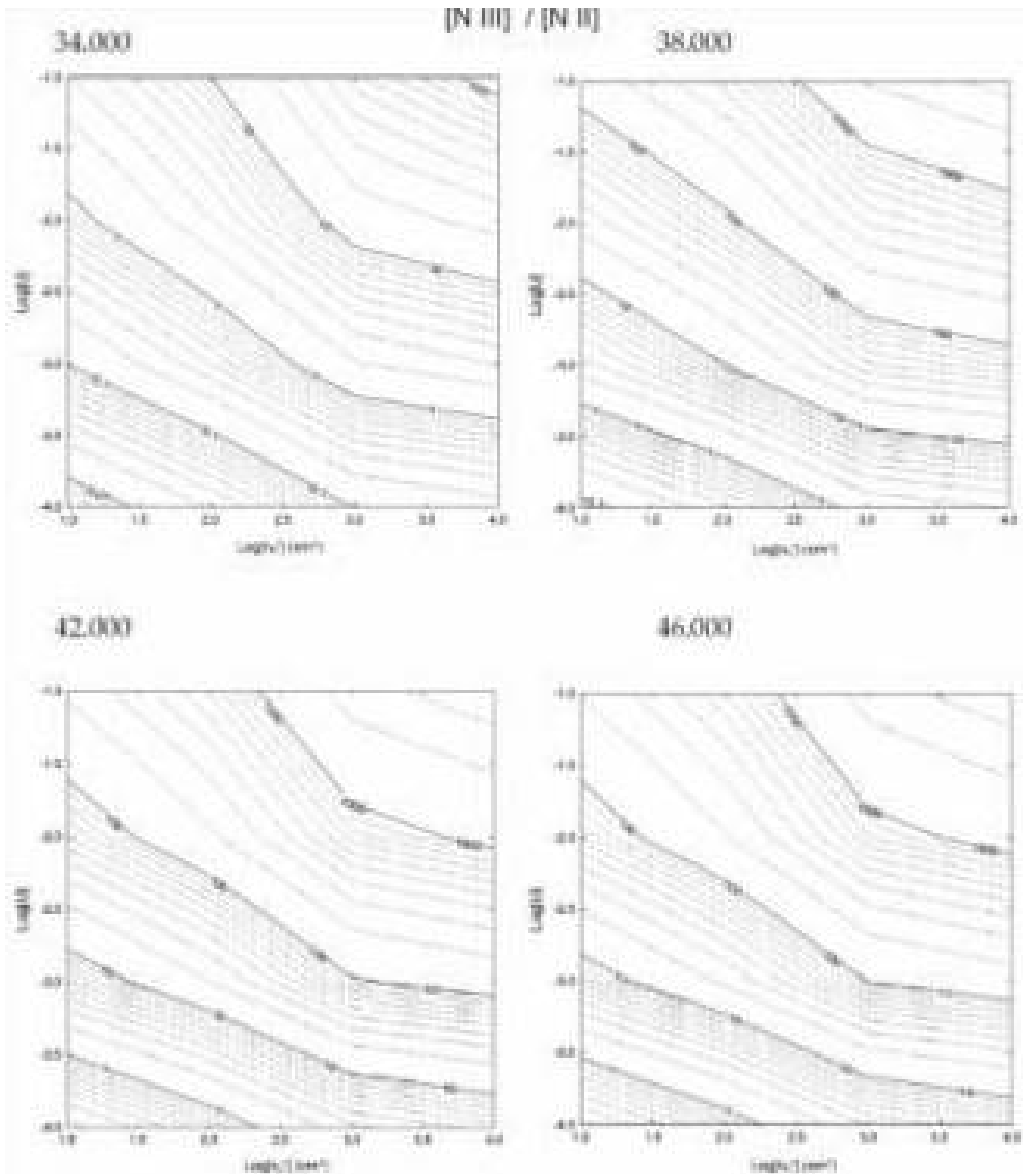

Figure 21 Ratio of the [N III] 57 μm intensity to the [N II] 205 μm intensity. The critical densities for these transitions are less than $10^4$ cm$^{-3}$ (Malhorta et al. 2001). This means that, for the range of $n_{H^+}$ considered in our calculations, this ratio can also depend on density in addition to $T$ and $U$.

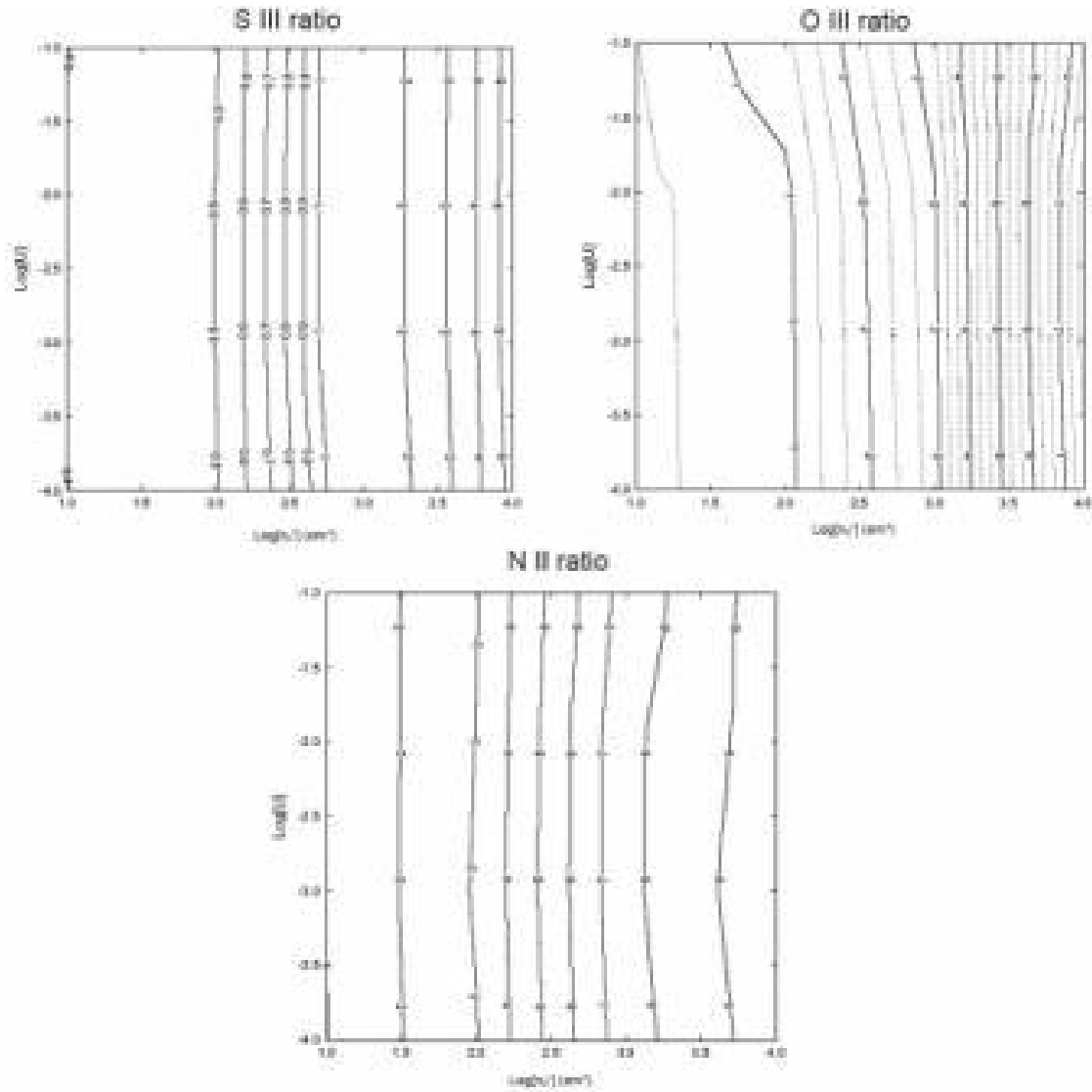

Figure 22 Computed H II region density diagnostic line ratios. Plotted is the ratio of the [S III] 18.7 μm line intensity to the [S III] 33.5 μm line, the [O III] 52 μm line intensity to the [O III] 88 μm line, and the [N II] 122 μm line intensity to the [N II] 205 μm line. All line ratios are good density diagnostics in the H II region for $n_{H^+} > 10^2$ cm$^{-3}$, while the N II ratio continues to be a good diagnostic across the entire range of $n_{H^+}$ considered.

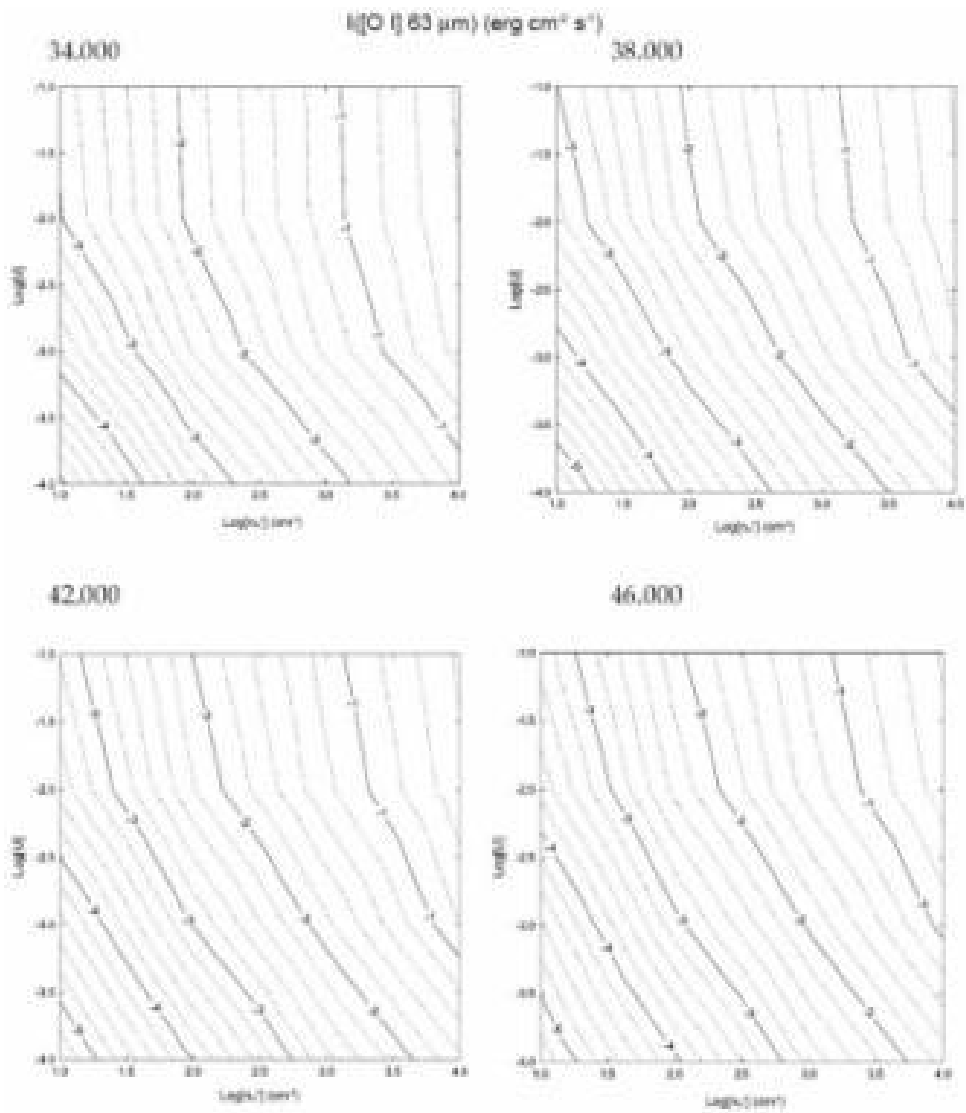

Figure 23 The predicted intensity of the [O I] 63 μm line (units ergs cm$^{-2}$ s$^{-1}$). This line and the lines presented in Figures 24-26 depend on both the hydrogen density in the PDR and $G_0$.

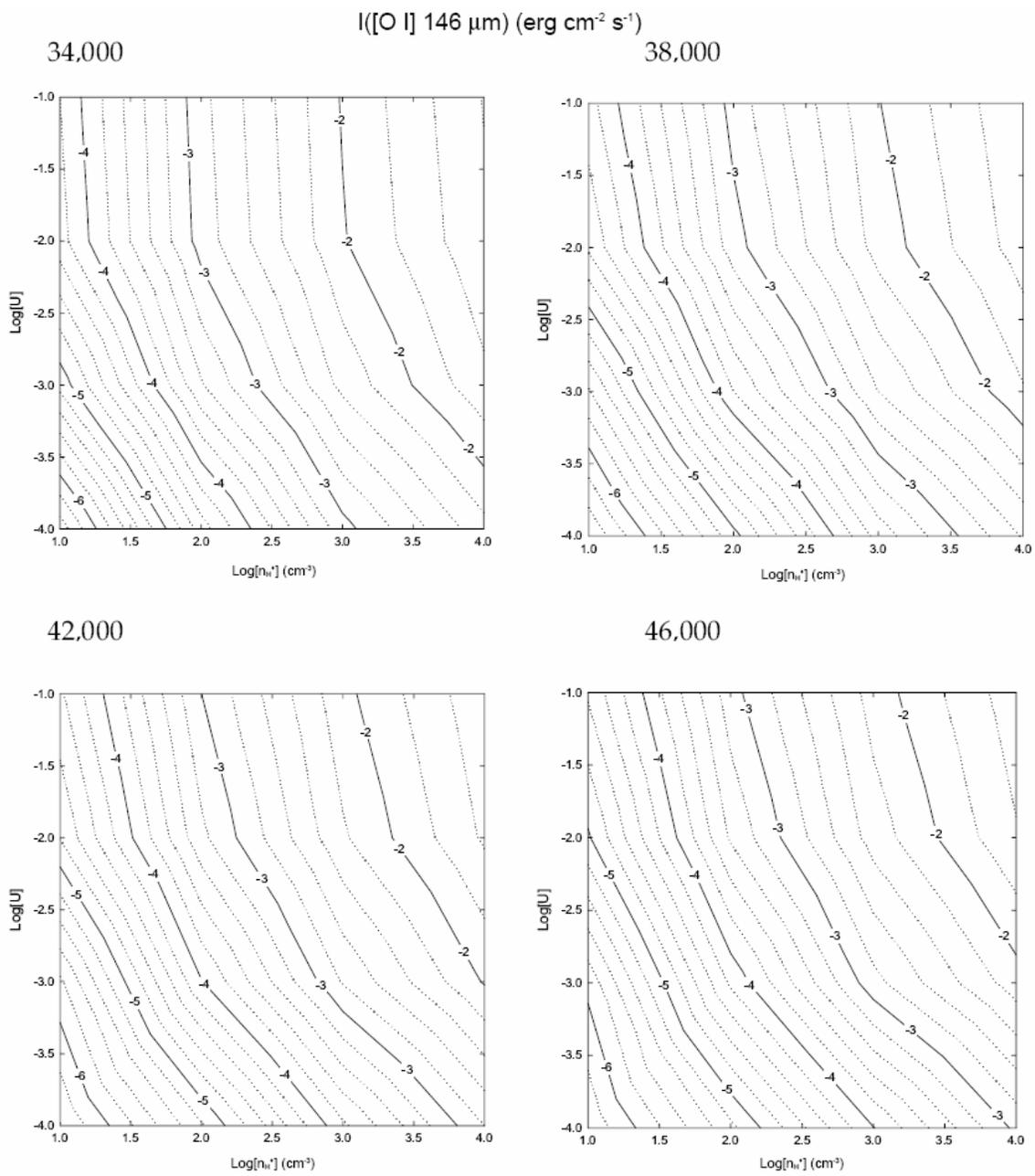

Figure 24 The predicted intensity of the [O I] 146 μm line (units of ergs cm$^{-2}$ s$^{-1}$).

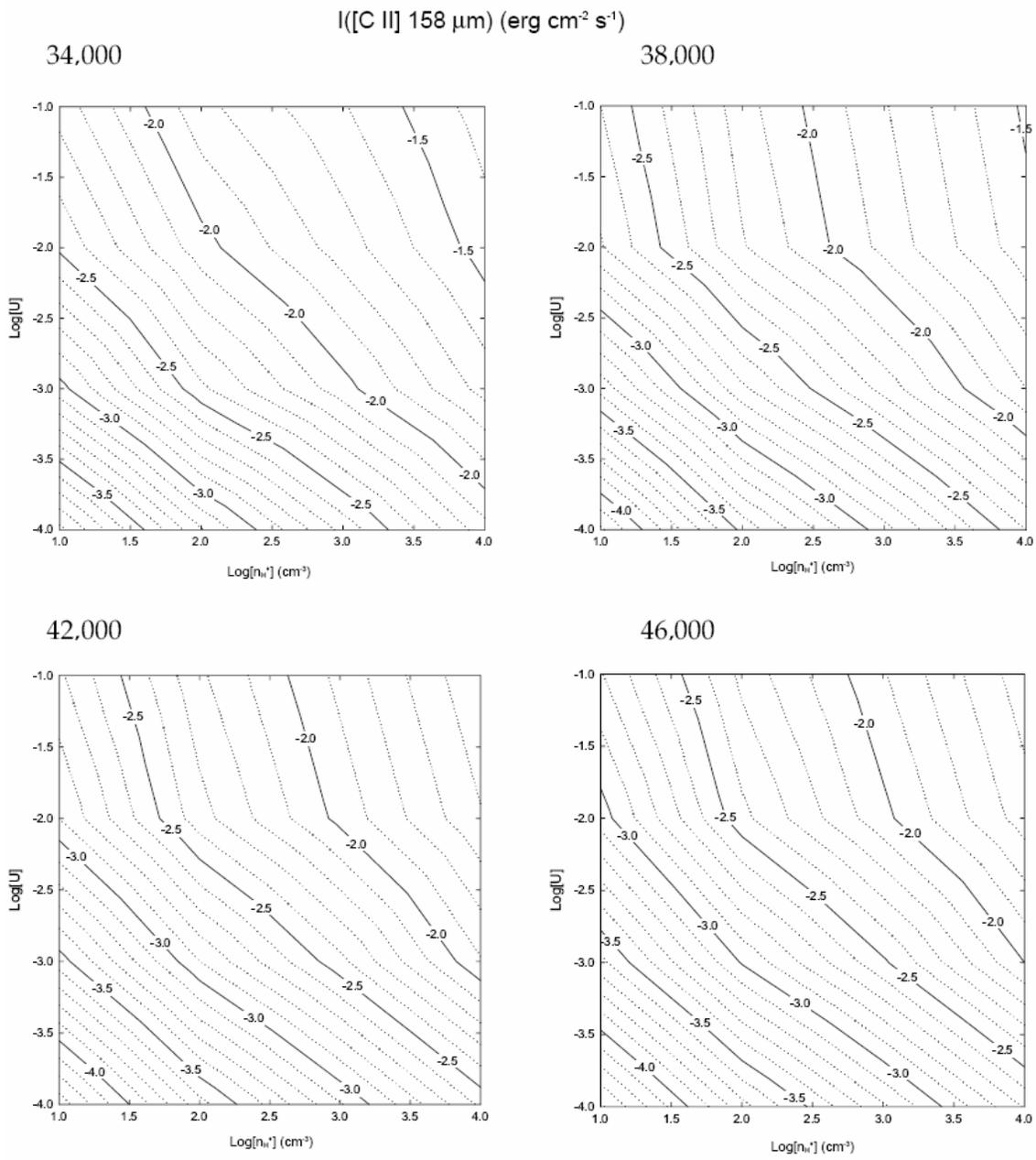

Figure 25 The predicted intensity of the [C II] 158 μm line (units of ergs cm$^{-2}$ s$^{-1}$).

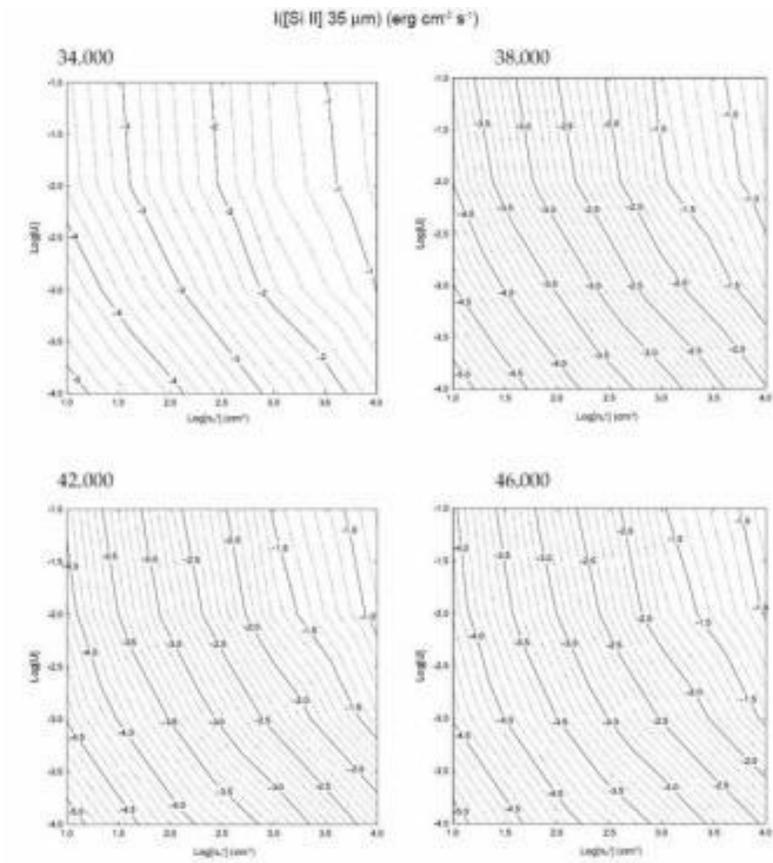

Figure 26 The predicted intensity of the [Si II] 35 µm line (units of ergs cm$^{-2}$ s$^{-1}$).

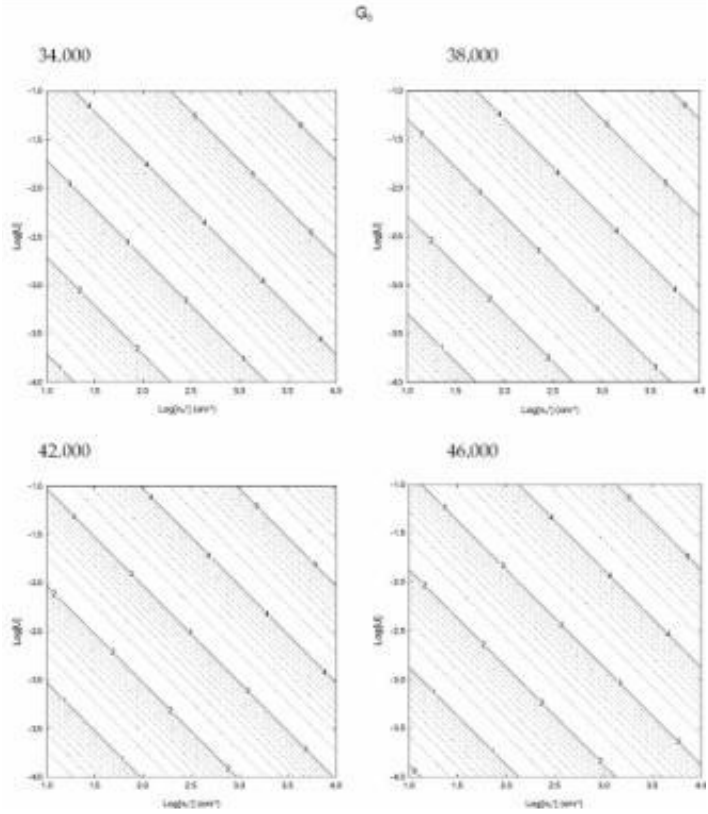

Figure 27 Log[$G_0$] at the illuminated face for our calculations. $G_0$ depends on $T^*$ and the cloud-star separation and can be posed into the product of $U$ and $n_H$ (see equation 1).

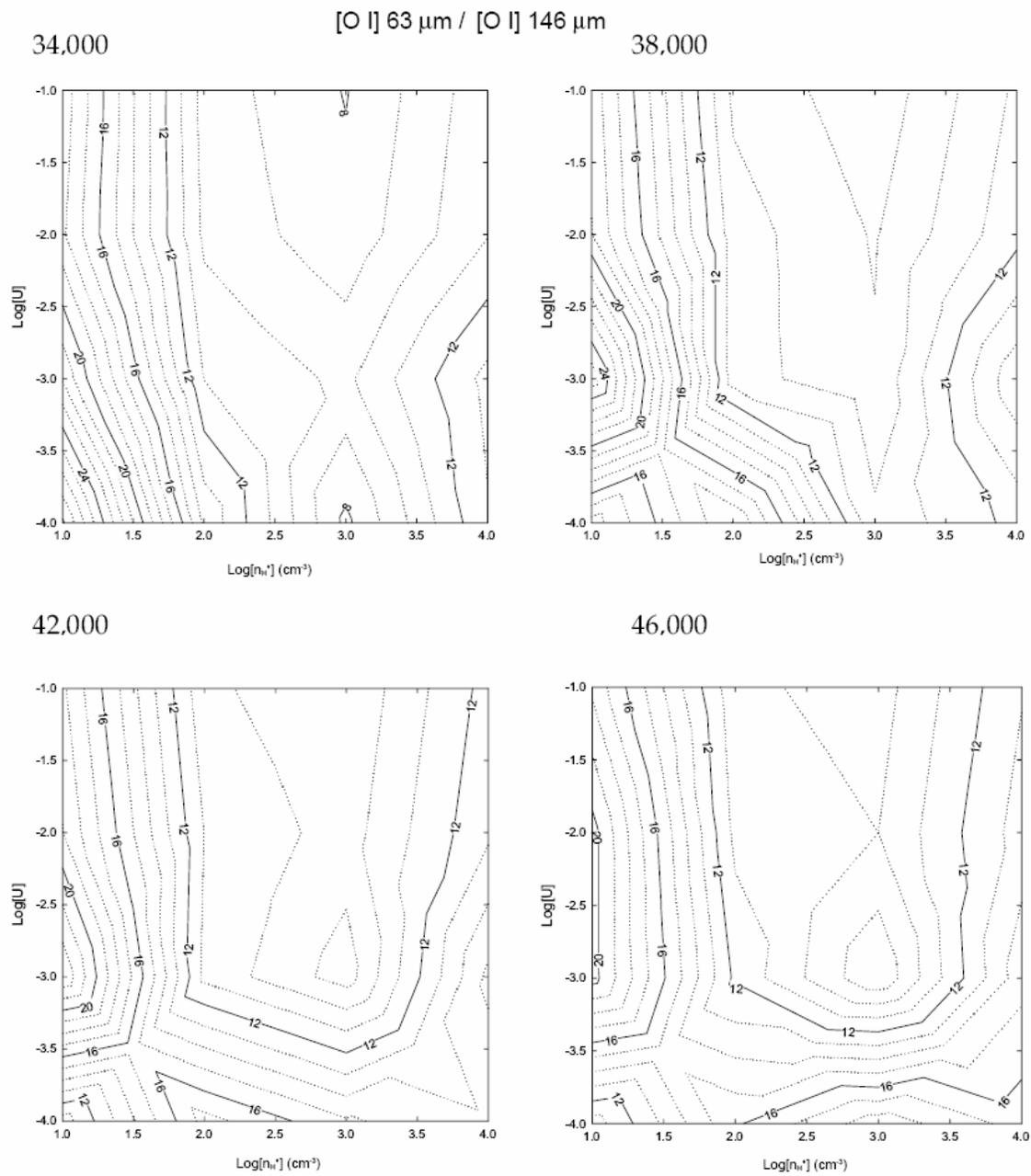

Figure 28  The predicted ratio of the [O I] 63 μm to [O I] 146 μm line intensities.  This ratio is a common diagnostic used to determine physical conditions in PDRs (Kaufman et al. 1999).  Over the entire range of parameter space, this ratio varies by only a factor of three.

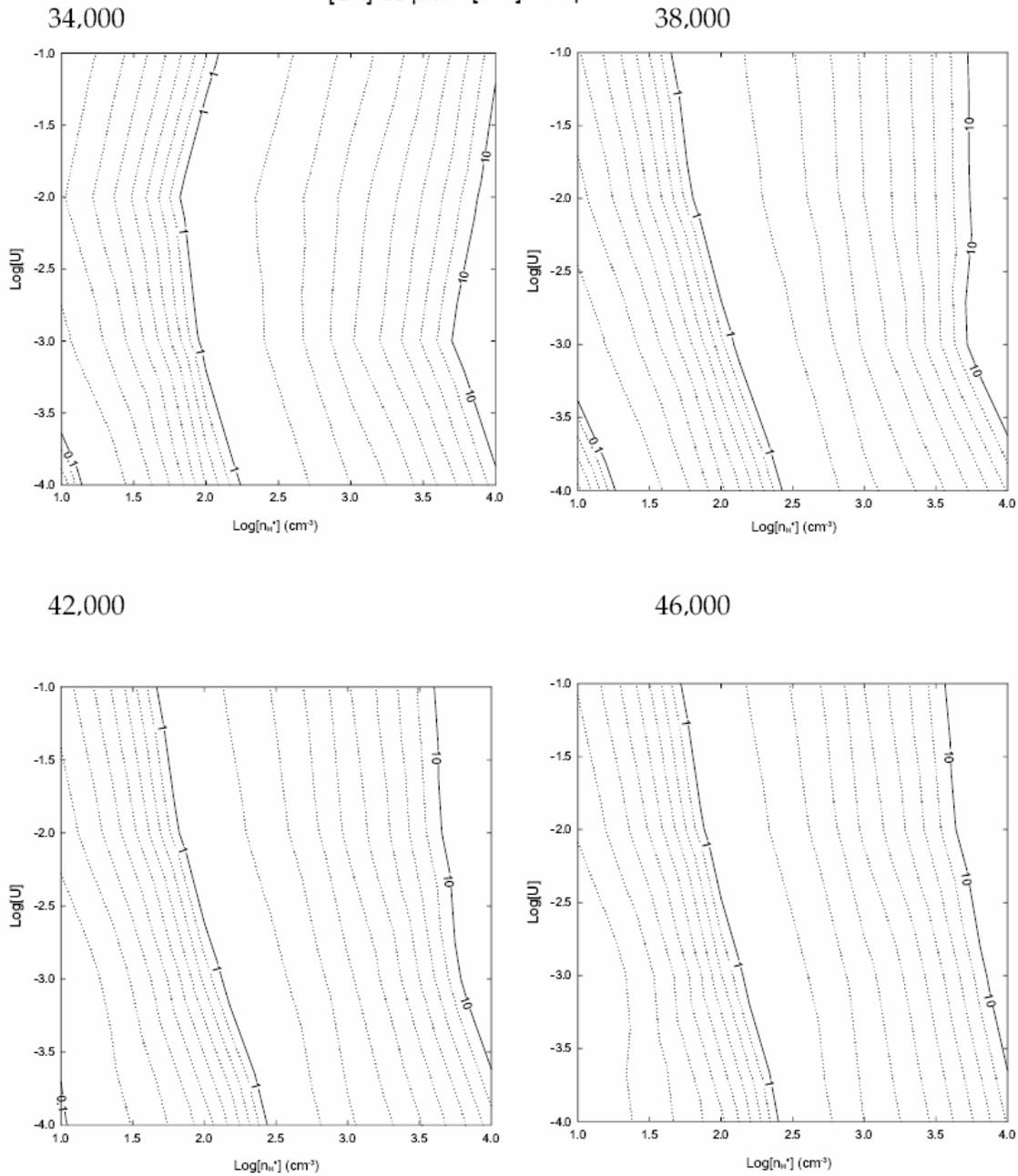

Figure 29 The predicted ratio of the [O I] 63 µm to [C II] 158 µm line intensities. This is another common PDR diagnostic line ratio (Kaufman et al. 1999). A significant amount of the 158 µm emission can come from the H II region. Without taking the H II region into account (see Figures 30-33), this ratio will be overpredicted, leading to errors in derived physical conditions that use this ratio.

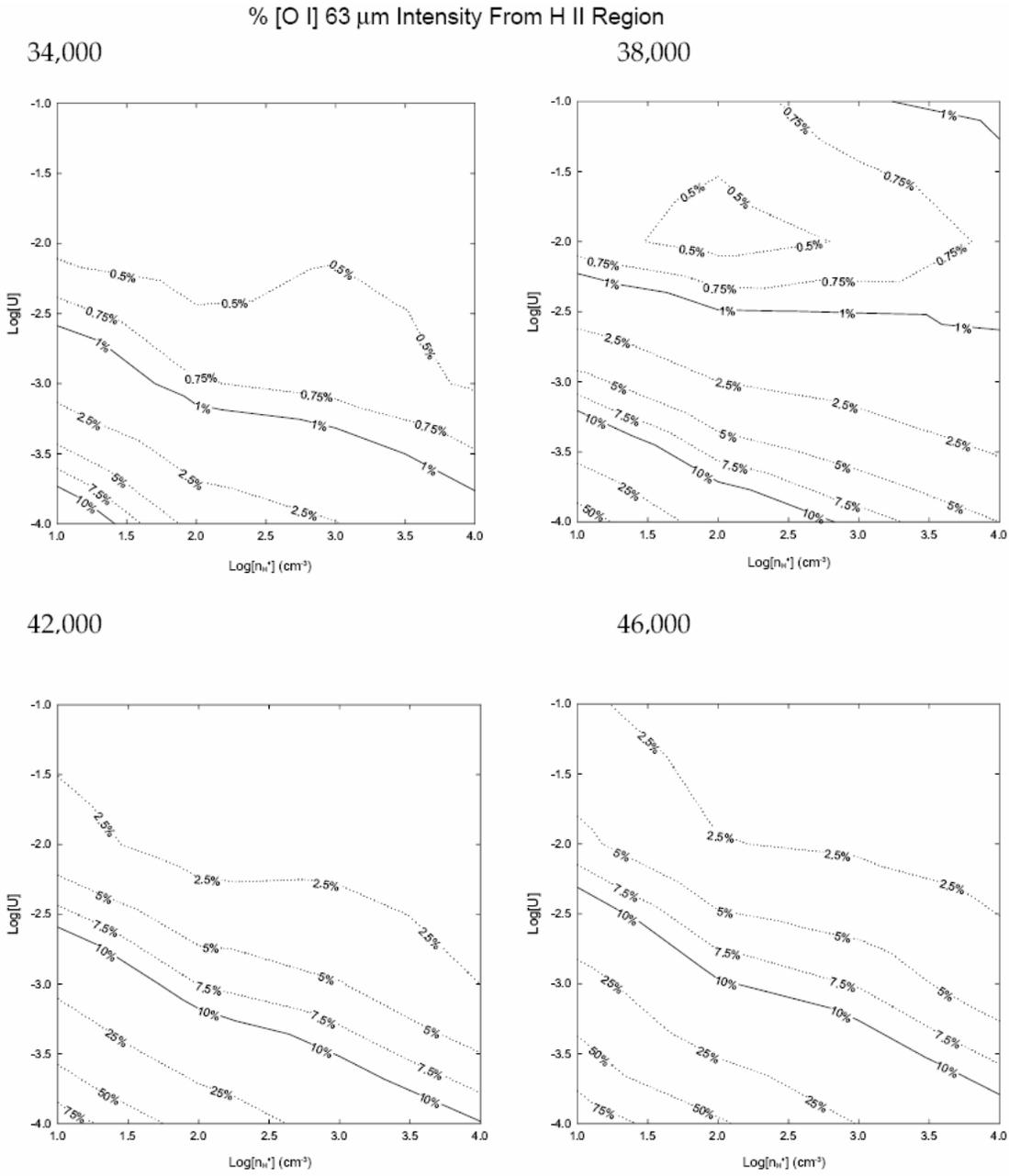

Figure 30 The percentage of the [O I] 63 μm line that emerges from the H II region for our calculations. The ionization structure of O follows that of H, so some $O^0$ exists in regions where hydrogen is partially ionized. For low $U \propto G_0/n_H$ environments, the size of the atomic region is small. This leads to a larger fraction of the total emission emerging from the H II region.

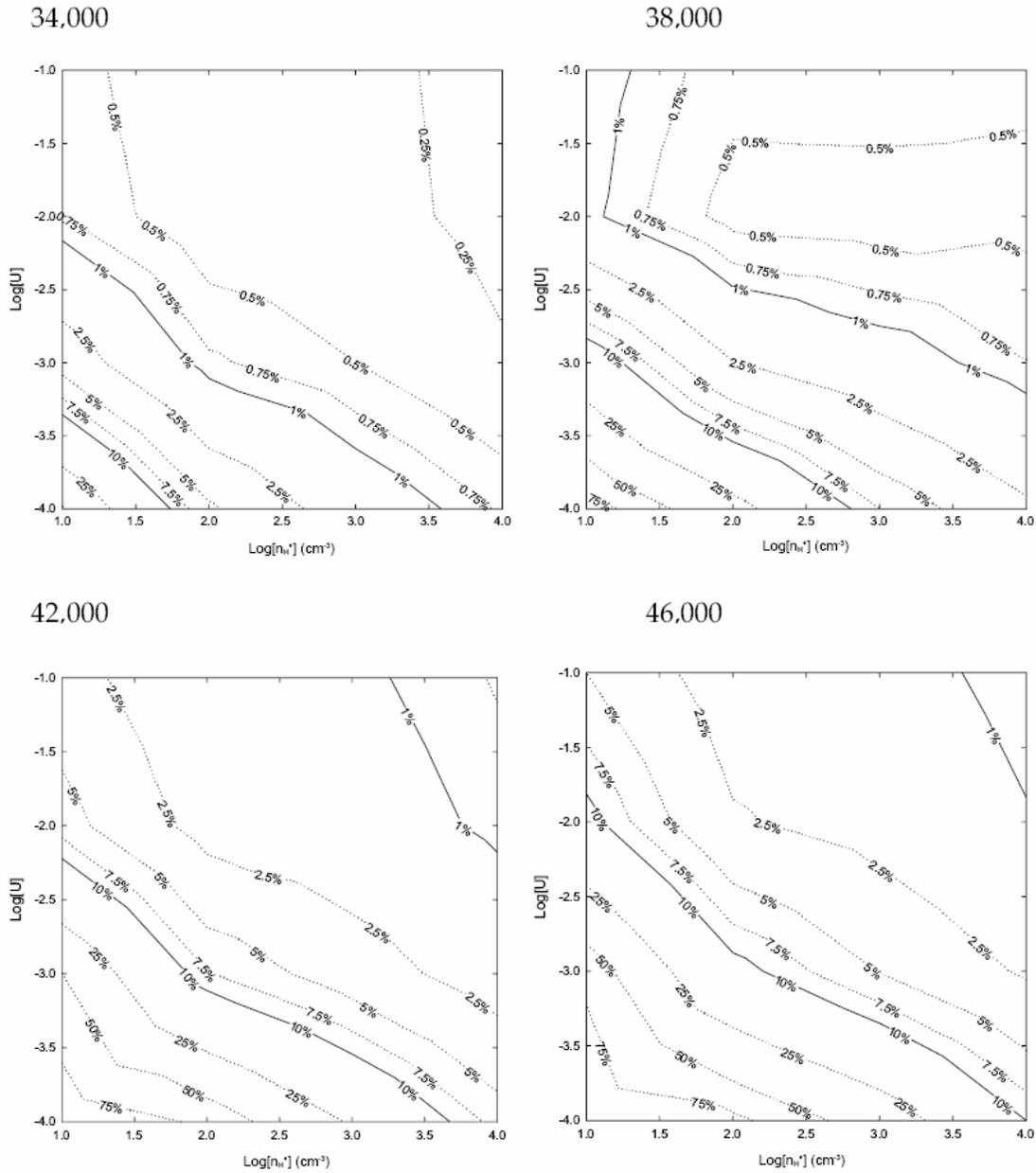

Figure 31  The percentage of the [O I] 146 µm line that is produced in the H II region.   The logic governing the H II region contribution to [O I] 146 µm emission is exactly the same as [O I] 63 µm.

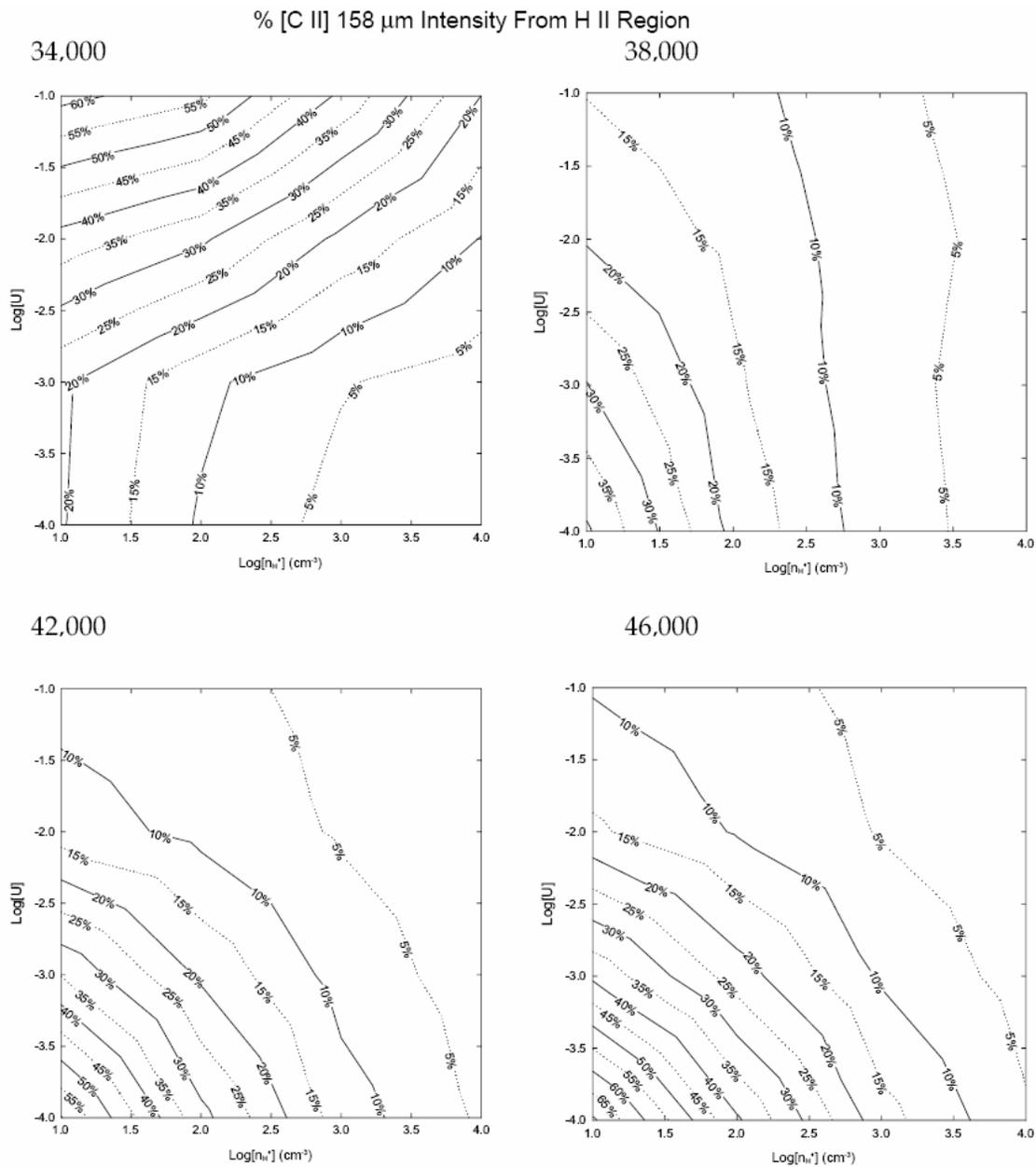

Figure 32 The percentage of the [C II] 158 μm line that is produced in the H II region. $C^0$ has an ionization potential less than 13.6 eV, which means that large amounts of $C^+$ can exist in ionized gas. For $T_* = 34,000$ K, $C^+$ and $C^{2+}$ are the dominant stages of ionization. This leads to a large H II region contribution to the 158 μm line. For higher $T_*$, more $C^{2+}$ and $C^{3+}$ is found in the H II region and its contribution to the total amount of C II emission is less.

% [Si II] 35 μm Intensity From H II Region

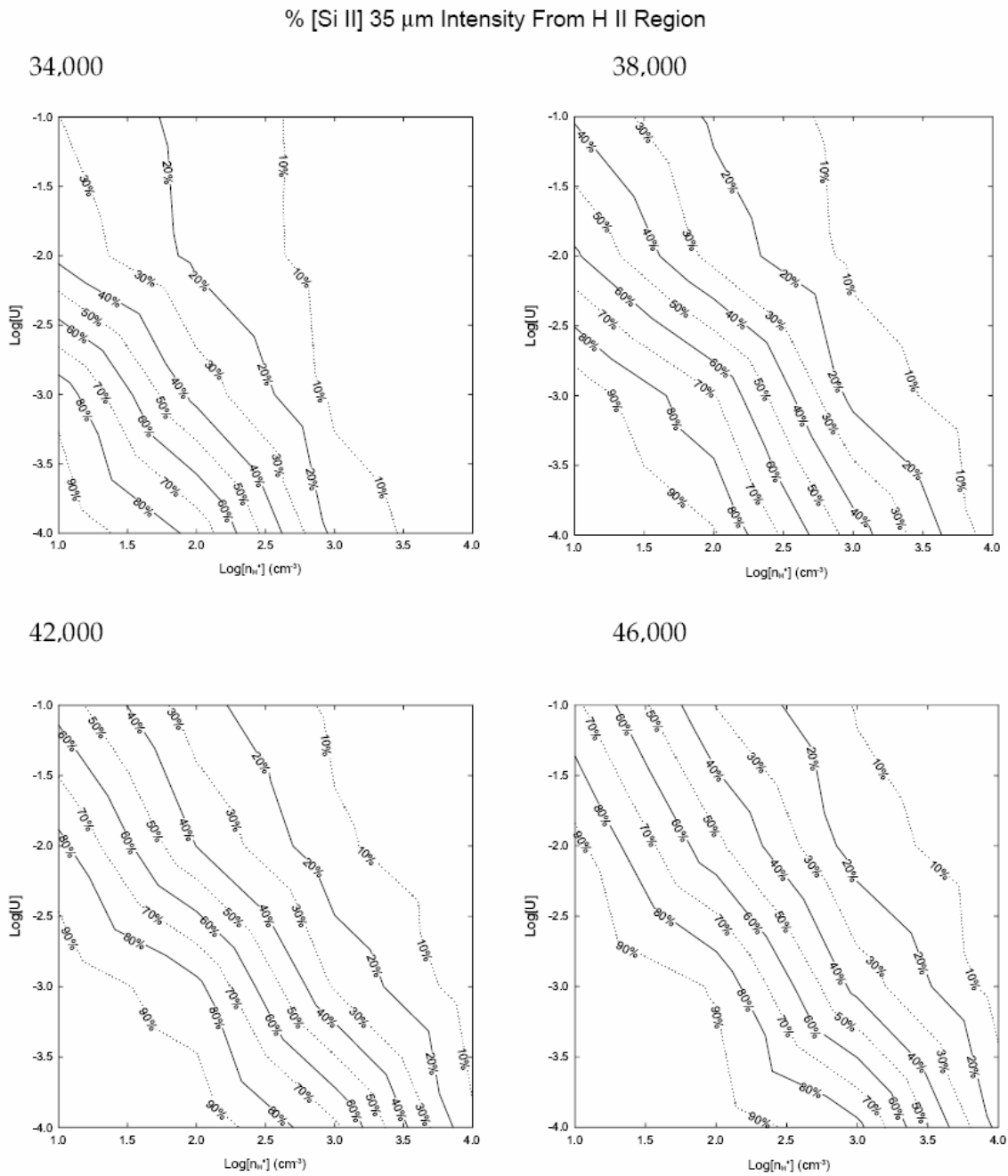

Figure 33  The percentage of the [Si II] 35 μm line that is produced in the H II region.  Silicon, like Carbon, has a first ionization potential < 13.6eV, which leads to a significant amount of the 35 μm line emission (>10%) coming from the H II region over the entire range of parameter space.

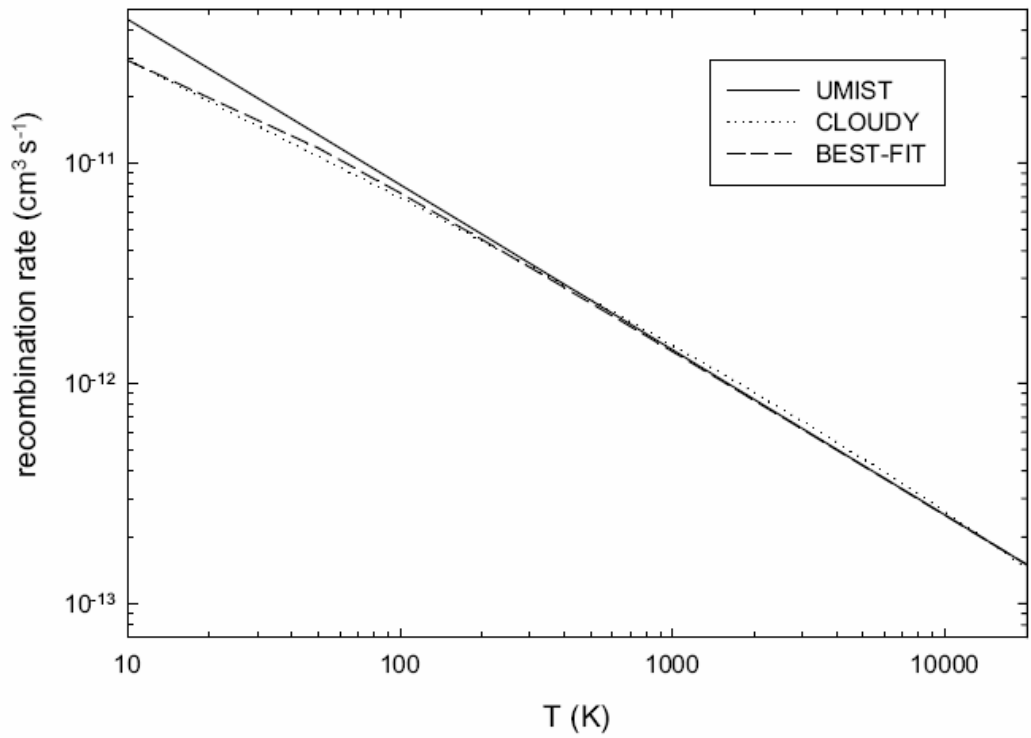

Figure 34 The hydrogen radiative recombination rate used in UMIST (solid line) and in Cloudy (dotted line) versus temperature. Also shown is the best-fit equation to the recombination rate (Ferland et al. 1992, dashed line).

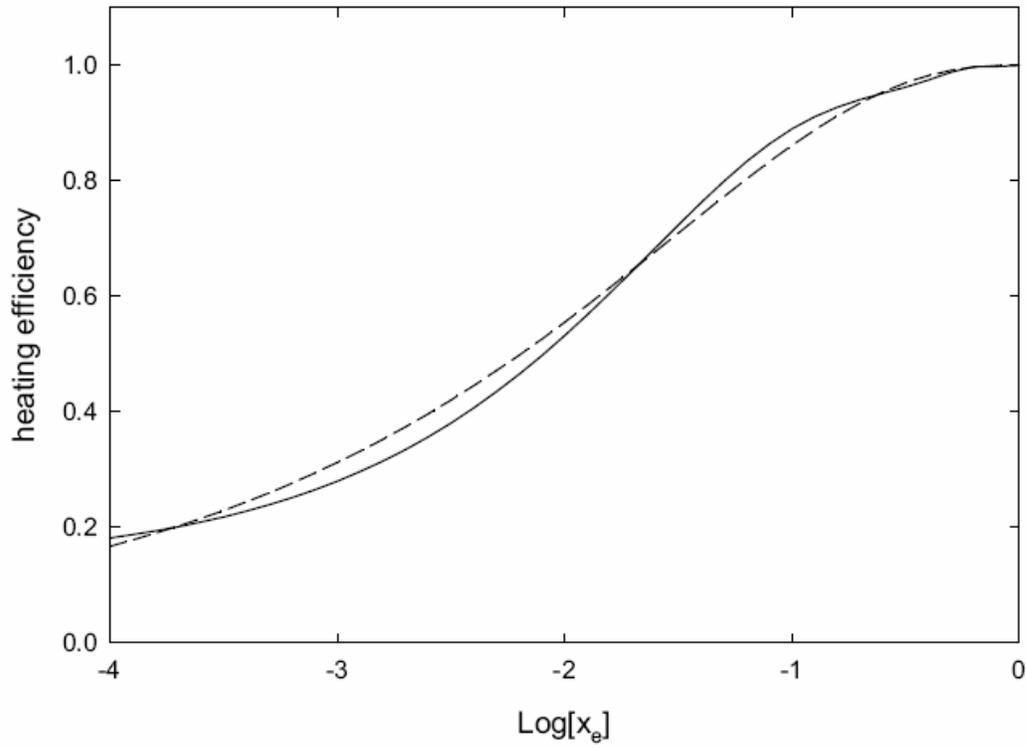

Figure 35 The cosmic ray heating efficiency variation with electron fraction. The solid line is the heating efficiency given by equation 2 of Appendix C, while the dashed line is equation A3 of Wolfire et al. (1995).